\documentclass[prd,preprint,superscriptaddress,showpacs,byrevtex]{revtex4}
\usepackage{bm}
\def\fsl#1{\setbox0=\hbox{$#1$}           
   \dimen0=\wd0                                 
   \setbox1=\hbox{/} \dimen1=\wd1               
   \ifdim\dimen0>\dimen1                        
      \rlap{\hbox to \dimen0{\hfil/\hfil}}      
      #1                                        
   \else                                        
      \rlap{\hbox to \dimen1{\hfil$#1$\hfil}}   
      /                                         
   \fi}                                         %
\newcommand{\be}{\begin{equation}}
\newcommand{\ee}{\end{equation}}
\newcommand{\bea}{\begin{eqnarray}}
\newcommand{\eea}{\end{eqnarray}}
\newcommand{\beq}{\begin{equation}}
\newcommand{\eeq}{\end{equation}}
\newcommand{\beqs}{\begin{eqnarray}}
\newcommand{\eeqs}{\end{eqnarray}}

\begin{document}
\title{ General Form of Color Charge of the Quark }
\author{Gouranga C Nayak } \email{nayak@max2.physics.sunysb.edu}
\affiliation{ 1001 East 9th Street, \#A, Tucson, AZ 85719, USA }
%
\begin{abstract}
In Maxwell theory the constant electric charge $e$ of the electron is consistent with the continuity
equation $\partial_\mu j^\mu(x)=0$ where $j^\mu(x)$ is the current density of the electron where
the repeated indices $\mu=0,1,2,3$ are summed. However,
in Yang-Mills theory the Yang-Mills color current density $j^{\mu a}(x)$ of the quark satisfies the equation
$D_\mu[A]j^{\mu a}(x)=0$ which is not a continuity equation ($\partial_\mu j^{\mu a}(x)\neq 0$)
which implies that the color charge of the quark is not constant where $a=1,2,...,8$ are the
color indices. Since the charge of a point particle is obtained from the zero ($\mu =0$) component of
a corresponding current density by integrating over the entire (physically) allowed volume, the
color charge $q^a(t)$ of the quark in Yang-Mills theory is time dependent. In this paper we
derive the general form of eight time dependent fundamental color charges $q^a(t)$ of the quark
in Yang-Mills theory in SU(3) where $a=1,2,...,8$.
\end{abstract}
\maketitle
\pagestyle{plain}
\pagenumbering{arabic}

\section{Introduction}

The electric and magnetic phenomena in the nature originates from the electric charge. The electric charge
of an electron is a fundamental quantity of the nature which is constant and has the experimentally measured
value $e=1.6\times 10^{-19}$ coulombs. Note that there are two types of electric charges in the nature,
1) positive (+) charge and 2) negative (-) charge. The charge of the electron is negative (-). The
electric charge produces the electromagnetic force in the nature.

However, inside a hadron (such as proton and neutron) the electromagnetic force can not bind the quarks
together. The force which is responsible for bound state hadron formation is called the strong force or the
color force which is a fundamental force of the nature. This strong force or the color force is not produced from
the electric charge of the quark but it is produced from the color charges of the quark. The color charge
is a fundamental charge of the nature which exists inside hadrons.

Note that in the Maxwell theory in electrodynamics the constant electric
charge $e$ of the electron is consistent with the continuity equation
\bea
\partial_\mu j^\mu(x) =0
\label{2c}
\eea
where $j^\mu(x)$ is the electric current density of the electron and $\mu=0,1,2,3$.
All the repeated indices are summed in this paper. However, in the Yang-Mills theory
the Yang-Mills color current density $j^{\mu a}(x)$ of the quark satisfies the equation \cite{yang,muta}
\bea
D_\mu[A]j^{\mu a}(x)=0,~~~~~~~~~~~~~~~~~~D_\mu^{ab}[A]=\delta^{ab}\partial_\mu + gf^{acb}A_\mu^c(x),~~~~~~~~a,b,c=1,2,...,8
\label{ccc}
\eea
which is not a continuity equation ($\partial^\mu j_\mu^a(x) \neq 0$)
which implies that the color charge of the quark is not constant where $A^{\mu a}(x)$ is
the Yang-Mills potential (color potential).
Since the charge of a point particle can be obtained from the zero ($\mu =0$) component of
a corresponding current density by integrating over the entire (physically) allowed volume, the
color charge $q^a(t)$ of the quark in Yang-Mills theory is time dependent.

Also, it is important that the conserved color
charges are not directly observable -- only color representations -- because
of the unbroken gauge invariance of QCD.
Thus, the concept of constant color charge seems unphysical.
The form of the color charge of the quark can be obtained from the zero component of the color current density
$j_0^a(x)$ of the quark. For earlier works on classical Yang-Mills theory, see \cite{wuyang,wuyang1,mandula,rest1,rest1a,rest1b,all}.

The color current density of the quark in eq. (\ref{ccc}) is related to the quark field $\psi_i(x)$ via
the equation \cite{yang,muta}
\bea
j^{\mu a}(x) = g {\bar \psi}_i(x) \gamma^\mu T^a_{ij} \psi_j(x),~~~~~~~~~~~~~~~~~~~a=1,2,...,8;~~~~~~~~~~~~~i,j=1,2,3
\label{dcurrentdss}
\eea
where $T^a=\frac{\lambda^a}{2}$ are eight generators of SU(3) and $\lambda^a$ are eight Gell-Mann matrices.
Since the color current density $j^{\mu a}(x)$ of a quark in eq. (\ref{dcurrentdss}) has eight color
components $a=1,2,...,8$ one finds that there are eight time dependent color charges of a quark.

We denote eight time dependent fundamental color charges of a quark by $q^a(t)$ where $a=1,2,...,8$ are color indices.
These eight time dependent fundamental color charges $q^a(t)$ of a quark are independent of quark flavor,
{\it i.e.}, a color charge $q^a(t)$ of the $u$ (up) quark is same as that of $d$ (down), $S$ (strange),
$c$ (charm), $B$ (bottom) or $t$ (top) quark.

It is useful to remember that the indices $i$=1,2,3=RED, BLUE, GREEN are not color charges of the
quark but they are color indices of the quark field $\psi_i(x)$ in eq. (\ref{dcurrentdss}). Color charges
$q^a(t)$ of a quark are functions
and they have values. This is analogous to electric charge '$-e$' of the electron which is not just a '$-$'
sign but it has a constant value $e=1.6\times 10^{-19}$ coulombs. Similarly RED, BLUE, GREEN symbols are not
color charges of a quark but $q^a(t)$ are the color charges of a quark where $a=1,2,..8$. Hence one finds that a
quark does not have three color charges but a quark has eight color charges. Another argument in favour of eight
color charges as opposed to three (RED, BLUE, GREEN) is that the latter depends on the representation while
the former, which is the number of generators, does not. In the Maxwell theory the electric
charge $e$ produces the electromagnetic force in the nature and in the Yang-Mills theory the color charges $q^a(t)$
produce the color force or the strong force in the nature.

In the Yang-Mills theory in SU(2) the color current density of a fermion is given by
\bea
j^{\mu i}(x) = g {\bar \psi}_k(x) \gamma^\mu \tau^i_{kn} \psi_n(x),~~~~~~~~~i=1,2,3;~~~~~~~~k,n=1,2
\label{su2ji}
\eea
where $\tau^i=\frac{\sigma^i}{2}$ are three generators of SU(2) and $\sigma^i$ are three Pauli matrices.
Hence one finds that there are three time dependent color charges $q_i(t)$ of a
fermion in the Yang-Mills theory in SU(2) where $i=1,2,3$. We find
that the general form of three time dependent color charges of a fermion in the Yang-Mills theory in SU(2) is
given by
\bea
&& q_1(t) = g\times {\rm sin}\theta(t)\times {\rm cos}\phi(t), \nonumber \\
&& q_2(t) = g\times {\rm sin}\theta(t)\times {\rm sin}\phi(t), \nonumber \\
&& q_3(t) = g\times {\rm cos}\theta(t)
\label{2skin}
\eea
where the time dependent real phases $\theta(t)$ and $\phi(t)$ can not be independent of time $t$ and the allowed ranges
of $\theta(t)$, $\phi(t)$ are given by
\bea
\frac{\pi}{3} \le \theta(t) \le \frac{2\pi}{3},~~~~~~~~~~~~-\pi < \phi(t) \le  \pi.
\label{2ina}
\eea
[See eqs. (\ref{2skfn}) and (\ref{2fna}) for the derivation of eqs. (\ref{2skin}) and (\ref{2ina})].

In this paper we will derive the general form of eight time dependent fundamental color charges
$q^a(t)$ of the quark in Yang-Mills theory in SU(3) where $a=1,2,...,8$.
We find that the general form of eight time dependent fundamental color
charges of the quark in Yang-Mills theory in SU(3) is given by
\bea
&& q_1(t) = g\times {\rm sin}\theta(t) \times {\rm sin}\sigma(t)\times {\rm cos}\eta(t)\times {\rm cos}\phi_{12}(t), \nonumber \\
&& q_2(t) = g\times {\rm sin}\theta(t) \times {\rm sin}\sigma(t)\times {\rm cos}\eta(t)\times {\rm sin}\phi_{12}(t), \nonumber \\
&& q_3(t) = g\times {\rm cos}\theta(t) \times {\rm sin}\phi(t) \nonumber \\
&& q_4(t) = g\times {\rm sin}\theta(t) \times {\rm sin}\sigma(t)\times {\rm sin}\eta(t)\times {\rm cos}\phi_{13}(t), \nonumber \\
&& q_5(t) = g\times {\rm sin}\theta(t) \times {\rm sin}\sigma(t)\times {\rm sin}\eta(t)\times {\rm sin}\phi_{13}(t), \nonumber \\
&& q_6(t) = g\times {\rm sin}\theta(t)\times {\rm cos}\sigma(t) \times {\rm cos}\phi_{23}(t), \nonumber \\
&& q_7(t) = g\times {\rm sin}\theta(t)\times {\rm cos}\sigma(t)\times {\rm sin}\phi_{23}(t), \nonumber \\
&& q_8(t) = g\times {\rm cos}\theta(t)\times {\rm cos}\phi(t)
\label{qspin}
\eea
where the ranges of the real time dependent phases are given by
\bea
&&{\rm sin}^{-1}(\sqrt{\frac{2}{3}}) ~~\le ~~ \theta(t) ~~ \le ~~\pi-{\rm sin}^{-1}(\sqrt{\frac{2}{3}}),~~~~~~~~~~~0 \le \sigma(t),~\eta(t) \le \frac{\pi}{2},\nonumber \\
&& 0~\le ~ \phi(t) \le 2 \pi,~~~~~~~~~~~~~~~~~~~~~~~~~~~~~~~~-\pi < \phi_{12}(t),~\phi_{13}(t),~\phi_{23}(t) \le \pi.
\label{3ina}
\eea

It can be seen that the general form of eight time dependent fundamental color charges $q^a(t)$
of the quark in eq. (\ref{qspin}) depend on the universal coupling $g$ which is a fundamental quantity
of the nature (a physical observable) that appears in the Yang-Mills Lagrangian density \cite{yang,muta}.

Since the fundamental time dependent color charge $q^a(t)$ of the quark
in eq. (\ref{qspin}) is linearly proportional to $g$ and the Yang-Mills potential
(color potential) $A^{\mu a}(x)$ contains infinite powers of $g$ (see eq. (\ref{fnab}) or \cite{arxiv})
we find that the definition of the fundamental
time dependent color charge $q^a(t)$ of the quark in eq. (\ref{qspin}) is independent of the Yang-Mills
potential $A^{\mu a}(x)$ [see section XIV for detailed discussion about this].

It can be seen that the general form of three time dependent color
charges $q_i(t)$ of a fermion in SU(2) in eq. (\ref{2skin})
and the general form of eight time dependent fundamental color charges
$q^a(t)$ of the quark in SU(3) in eq. (\ref{qspin}) are consistent with the fact that SU(2) and SO(3) are locally
isomorphic, while SU(3) and SO(8) are not [see sections IX and XV for more discussion on this].

It should be remembered that the universal coupling $g$
(which is a physical observable, a fundamental quantity of the nature)
is the only parameter (apart from the mass $m$ of the quark)
that appears in the classical Yang-Mills lagrangian density
\cite{yang,muta} (see eq. (\ref{yml}) below). Note that in the derivation of
eqs. (\ref{qspin}) and (\ref{3ina}) we have fixed the first Casimir (quadratic Casimir)
invariant
\bea
q^a(t)q^a(t)=g^2
\label{fcinv}
\eea
of SU(3) to be $g^2$ [see eqs. (\ref{fly}) and (\ref{qb2shj})]. Hence we find that
the general form of eight time dependent fundamental color charges
$q^a(t)$ of the quark in Yang-Mills theory in SU(3) in eq. (\ref{qspin})
depend on $g$ and seven time dependent phases
$\theta(t),~ \sigma(t),~\eta(t),~ \phi(t),~ \phi_{12}(t),~\phi_{13}(t),~\phi_{23}(t)$
where the ranges of these seven time dependent phases are given by eq. (\ref{3ina}).
Since the first Casimir (quadratic Casimir) invariant $q^a(t)q^a(t)$
and the second Casimir (cubic Casimir) invariant $d_{abc}q^a(t)q^b(t)q^c(t)$ of SU(3) are two
independent Casimir invariants, one expects that if the second Casimir (cubic Casimir) invariant
$d_{abc}q^a(t)q^b(t)q^c(t)$ of SU(3)
corresponds to any physical observable then that physical observable should be experimentally measured
and that physical observable should be different from $g$ because the first Casimir (quadratic Casimir)
invariant $q^a(t)q^a(t)$ of SU(3) is fixed to be $g^2$, see eq. (\ref{fcinv}).
If such a physical observable exists in the nature and is fixed to be, say $C_3$,
[for example by experiments] where the fixed $C_3$ is given by
\bea
d_{abc}q^a(t)q^b(t)q^c(t)=C_3
\label{3fix}
\eea
then one finds that
\bea
&&\phi_{13}(t)~=~\phi_{12}(t)+\phi_{23}(t)~+~{\rm cos}^{-1}[\frac{1}{{\rm sin}^3\theta(t) ~ {\rm sin^2}\sigma(t)~{\rm cos}\sigma(t)~ {\rm sin}2\eta(t)}\times [\frac{2C_3}{3g^3}\nonumber \\
&&-~ {\rm sin}^2\theta(t)~ {\rm cos}\theta(t) ~ {\rm sin}\phi(t) ~ [{\rm sin}^2\sigma(t)~ {\rm sin}^2\eta(t)- {\rm cos}^2\sigma(t)]\nonumber \\
&&-\frac{1}{\sqrt{3}} ~{\rm cos}\theta(t) ~{\rm cos}\phi(t)~ [3~{\rm sin}^2\theta(t)~ {\rm sin}^2\sigma(t) ~{\rm cos}^2\eta(t)+3~{\rm cos}^2\theta(t)~ {\rm sin}^2\phi(t)+\frac{{\rm cos}^2\theta(t)~{\rm cos}^2\phi(t)}{3}-1]]]\nonumber \\
\label{phi13in}
\eea
[see eq. (\ref{phi13}) for derivation of eq. (\ref{phi13in})] in which case the general form of eight time dependent fundamental
color charges $q^a(t)$ of the quark in eq. (\ref{qspin}) depend on $g$, $C_3$ and six time dependent phases
$\theta(t),~ \sigma(t),~\eta(t),~ \phi(t),~ \phi_{12}(t),~\phi_{23}(t)$ where $\phi_{13}(t)$ is
given by eq. (\ref{phi13in}). However, note that even if all the physical observables are gauge invariant
but not all the gauge invariants are physical observables. Hence if there exists no physical
observable in the nature which is related to the fixed value $C_3$ as given by eq. (\ref{3fix})
[for example if one can not find any such observable from the experiments] then the second Casimir invariant
(cubic Casimir invariant) $d_{abc}q^a(t)q^b(t)q^c(t)$ of SU(3) satisfies the range
$-\frac{g^3}{\sqrt{3}}\le d_{abc}q^a(t)q^b(t)q^c(t) \le \frac{g^3}{\sqrt{3}}$
[see eq. (\ref{dabc})] in which case the general form of eight time dependent fundamental
color charges $q^a(t)$ of the quark in eq. (\ref{qspin}) depend on $g$ and seven time dependent phases
$\theta(t),~ \sigma(t),~\eta(t),~ \phi(t),~ \phi_{12}(t),~\phi_{13}(t),~\phi_{23}(t)$
where the ranges of these seven time dependent phases are given by eq. (\ref{3ina}) [see section XV for
detailed discussion on this].

Hence we find that the general form of eight time dependent fundamental color charges of the quark
in Yang-Mills theory in SU(3) is given by eq. (\ref{qspin}) where
$\theta(t),~\sigma(t),~\eta(t),~\phi(t),~\phi_{12}(t),~\phi_{13}(t),~\phi_{23}(t)$
are real time dependent phases. Note that if all of these seven phases become constants then
all the eight color charges $q^a$ become constants in which case
the Yang-Mills potential $A^{\mu a}(x)$ reduces to Maxwell-like (abelian-like) potential
(see eqs. (\ref{fnab}) and (\ref{upg1fg}) or \cite{arxiv}). Since the abelian-like potential can not
explain confinement of quarks inside (stable) proton one finds that all these seven real phases
can not be constants. Hence one finds that
the general form of eight time dependent fundamental color charges $q^a(t)$ of the quark which we have
derived in eq. (\ref{qspin}) may provide an insight to the question why
quarks are confined inside a (stable) proton once the exact form of these time dependent phases
$\theta(t),~ \sigma(t),~\eta(t),~ \phi(t),~ \phi_{12}(t),~\phi_{13}(t),~\phi_{23}(t)$ are found out
[see section XVI for a detailed discussion on this]. It should be remembered that the
static systems in the Yang-Mills theory are, in general, not
abelian-like. For example, unlike Maxwell theory where the electron at rest produces
(abelian) Coulomb potential, the quark at rest in Yang-Mills theory does not
produce abelian-like potential (see eq. (\ref{upg1fg}) and sections VII, VIII or \cite{arxiv}
for details).

We will provide the derivations of eqs. (\ref{qspin}) and (\ref{3ina}) in this paper .

The paper is organized as follows. In section II we discuss the expression of the abelian
pure gauge potential produced by the electron in Maxwell theory. In section III we
discuss the charge and the charge density of a point particle using quantum mechanics.
In section IV we show that the color charge of the quark in Yang-Mills theory is time
dependent. The relation between the coupling constant and the color charge
in Yang-Mills theory is discussed in section V. In section VI we discuss the color
current density, the color charge of the quark and the general form of the Yang-Mills potential.
In section VII we discuss the general form of the Yang-Mills potential (color potential) produced by
the quark at rest.
In section VIII we discuss the Yang-Mills color current density of the quark at rest.
In section IX we describe the analogy between Maxwell theory and Yang-Mills theory to
obtain the form of the fundamental charge of the fermion from the Dirac wave function.
In section X we discuss the fermion
color current density, the fermion wave function and the Pauli Matrices in Yang-Mills theory in SU(2). In
section XI we derive the general form of three time dependent fundamental color charges $q_i(t)$ of a
fermion in Yang-Mills theory in SU(2) where $i=1,2,3$. In section XII we discuss the Yang-Mills color
current density of the quark, the quark wave function and the Gell-Mann Matrices in Yang-Mills theory in SU(3). In
section XIII we derive the general form of eight time dependent fundamental color charges $q^a(t)$ of
the quark in Yang-Mills theory in SU(3) where $a=1,2,...,8$. In section XIV we compare our investigation
with the Wong's equation. In section XV we show that the general form of eight time dependent color
charges $q^a(t)$ of the quark is consistent with the fact that there is second (cubic) casimir invariant of
SU(3). The advantage of the time dependent phases in the color charge of
the quark is discussed in section XVI. Section XVII contains conclusion.

\section{ Abelian Pure Gauge Potential in Maxwell Theory }

The Maxwell equation in electrodynamics is given by \cite{jackson}
\bea
\partial_\nu F^{\nu \mu} =j^\mu,~~~~~~~~~~~~~~~~~~~~~~~~~~~~\partial_\mu F_{\nu \beta}+ \partial_\nu F_{\beta \mu }+ \partial_\beta F_{\mu \nu} =0
\label{maxwell}
\eea
where
\bea
F_{\mu \nu} =\partial_\mu A_\nu -\partial_\nu A_\mu.
\label{fmn}
\eea
In classical electrodynamics if the electric charge $e$ is of a point particle whose position in the
inertial frame $K$ is ${\vec X}(t)$ then the charge density $\rho(t,{\vec x})$ and current
density ${\vec j}(t,{\vec x})$ of the point charge $e$ in that frame are given by \cite{jackson}
\bea
&&\rho(t,{\vec x})= j_0(t,{\vec x}) = e~\delta^{(3)}({\vec x}-{\vec X}(t)), \nonumber \\
&& {\vec j}(t,{\vec x}) = e~{\vec v}(t)~\delta^{(3)}({\vec x}-{\vec X}(t))
\label{kcurrent}
\eea
where
\bea
{\vec v}(t) =\frac{d{\vec X}(t)}{dt}
\eea
is the charge's velocity in that frame $K$. From eq. (\ref{kcurrent}) one finds
\bea
&& \frac{\partial j_0(t,{\vec x})}{\partial t} = -e~v_x(t)~\delta'(x-{ X}(t)) ~\delta({ y}-{ Y}(t))~\delta({ z}-{ Z}(t)) \nonumber \\
&&-e~v_y(t)~\delta(x-{ X}(t)) ~\delta'({ y}-{Y}(t))~\delta({ z}-{ Z}(t))-e~v_z(t)~\delta(x-{ X}(t)) ~\delta({ y}-{ Y}(t))~\delta'({ z}-{ Z}(t))\nonumber \\
\label{kcurrent1}
\eea
where
\bea
\delta'(w) = \frac{d[\delta(w)]}{dw}
\eea
and
\bea
&& {\vec \nabla} \cdot {\vec j}(t,{\vec x}) = e~v_x(t)~\delta'(x-{ X}(t)) ~\delta({ y}-{ Y}(t))~\delta({ z}-{ Z}(t)) \nonumber \\
&&+e~v_y(t)~\delta(x-{ X}(t)) ~\delta'({ y}-{Y}(t))~\delta({ z}-{ Z}(t))+e~v_z(t)~\delta(x-{ X}(t)) ~\delta({ y}-{ Y}(t))~\delta'({ z}-{ Z}(t)).\nonumber \\
\label{kcurrent2}
\eea
From eqs. (\ref{kcurrent1}) and (\ref{kcurrent2}) one finds
\bea
\partial_\mu j^\mu(x)=\frac{\partial j^\mu(x)}{\partial x^\mu}= \frac{\partial j_0(t,{\vec x})}{\partial t} + {\vec \nabla} \cdot {\vec j}(t,{\vec x}) =0
\label{concurrent1}
\eea
which is the continuity equation in Maxwell theory. Hence one finds that the constant electric charge $e$ of the electron
satisfies the continuity eq. (\ref{concurrent1}).

In the covariant formulation the current density of the electron of charge $e$ is given by \cite{jackson}
\bea
j^{\mu}(x) =  \int d\tau ~e~u^\mu(\tau)~\delta^{(4)}(x-X(\tau))
\label{amx9}
\eea
which satisfies the continuity equation
\bea
\partial_\mu j^\mu(x) =0
\label{bmx9}
\eea
where
\bea
u^\mu(\tau) = \frac{dX^\mu(\tau)}{d\tau}
\eea
is the four-velocity of the electron and $X^\mu(\tau)$ four-coordinate of the electron.

The solution of the inhomogeneous wave equation
\bea
\partial_\nu \partial^\nu A^\mu(x)=j^\mu(x)
\label{amx2v}
\eea
is given by
\bea
A^\mu(x)=\int d^4x' D_r(x-x')j^\mu(x')
\label{bmx3}
\eea
where $D_r(x-x')$ is the retarded Greens function \cite{jackson}. From
eqs. (\ref{amx9}) and (\ref{bmx3}) we find that the electromagnetic potential
(Lienard-Weichert potential) produced by the constant electric charge $e$ of
the electron is given by
\bea
A^\mu(x) =e\frac{u^\mu(\tau_0)}{u(\tau_0) \cdot (x-X(\tau_0))}
\label{amx3}
\eea
which satisfies the Lorentz gauge condition
\bea
\partial_\mu A^\mu(x) =0
\label{amxl}
\eea
where $\tau_0$ is obtained from the solution of the retarded equation
\bea
x_0 -X_0(\tau_0) = |{\vec x}-{\vec X}(\tau_0)|.
\label{mx11}
\eea
In eq. (\ref{amx3}) the four-vector $x^\mu$ is the time-space position at which
the electromagnetic field is observed and the four-vector $X^\mu(\tau_0)$ is the
time-space position of the electron where the electromagnetic field was produced.

Hence we find that if $A^\mu(x)$ satisfies Lorentz gauge condition $\partial_\mu A^\mu(x) =0$
then eq. (\ref{amx2v}) reduces to the inhomogeneous Maxwell equation
\bea
\partial_\nu F^{\nu \mu }(x)=j^\mu(x)
\label{cmx1}
\eea
where
\bea
F^{\mu \nu}(x) =\partial^\mu A^\nu(x) -\partial^\nu A^\mu(x).
\label{aafmn}
\eea
Note that an electron has non-zero mass (even if it is very small) which implies that an electron can not travel
exactly at speed of light $v=c$. Since electromagnetic wave travels exactly at speed of light and the electron
can not travel exactly at speed of light one finds that at a common time $t=\frac{x_0}{c}=\frac{X_0(\tau)}{c}$
the observation point of the electromagnetic potential (or any gauge invariant obtained from the electromagnetic
potential) is given by ${\vec x} \neq {\vec X}(\tau)$ which implies from eq. (\ref{amx3}) and (\ref{aafmn}) that
\bea
\partial_\nu F^{\nu \mu}(x)= 0
\label{amxka}
\eea
which satisfies Maxwell equation given by eq. (\ref{cmx1}) where $j^\mu(x)$ is given by eq. (\ref{amx9}).

When the electron in uniform motion is at its highest speed
(which is arbitrarily close to the speed of light $v\sim c$) we find from eq. (\ref{amx3})
\bea
A^\mu(x)=e\frac{\beta^\mu_{\sim c}}{\beta_{\sim c} \cdot (x-X(\tau_0))},~~~~~~~~~~\beta^\mu_{\sim c}=(1,~{\vec \beta}_{\sim c}),~~~~~~~~~~~~{\vec \beta}^2_{\sim c} =\frac{v^2}{c^2} \sim 1
\label{amxk}
\eea
where $u^\mu=c \gamma \beta^\mu$ with $\gamma=\frac{1}{\sqrt{1-{\vec \beta}^2}}=\frac{1}{\sqrt{1-\frac{{ v}^2}{c^2}}}$.
From eq. (\ref{aafmn}) we find that $A^\mu(x)$ in eq. (\ref{amxk}) produced by the electron in uniform motion
at its highest speed (which is arbitrarily close to speed of light $v\sim c$) gives \cite{sterman,arxiv}
\bea
F^{\mu \nu}(x)\sim 0,
\label{amxq}
\eea
at all the time-space observation points $x^\mu$ (except at the spatial position ${\vec x}$ transverse to
the motion of the electron at the time of closest approach), see section 3.1 of \cite{arxiv} for details.
From eqs. (\ref{amxq}) and (\ref{aafmn}) we find that at all the time-space points $x^\mu$
(except at the spatial position ${\vec x}$ transverse to the motion of the electron at the time of closest
approach) the $A^\mu(x)$ in eq. (\ref{amxk}) can be written in the form \cite{sterman,arxiv}
\bea
A^\mu =e\frac{\beta^\mu_{\sim c}}{\beta_{\sim c} \cdot (x-X(\tau_0))} \sim \partial^\mu \omega(x),~~~~~~~~~~~{\rm where}~~~~~~~~~~\omega(x) = e~{\rm ln}[\beta_{\sim c} \cdot (x-X(\tau_0))].\nonumber \\
\label{amxs}
\eea
From eq. (\ref{amxs}) we find that the electron in uniform motion at its highest speed
(which is arbitrarily close to the speed of light $v\sim c$) produces U(1) (approximate) pure gauge potential
or abelian (approximate) pure gauge potential $A^\mu \sim \partial^\mu \omega(x)$ at all the time-space points $x^\mu$
(except at the spatial position ${\vec x}$ transverse to the motion of the electron at the time of closest
approach). We call the electromagnetic potential $A^\mu(x)$ in eq. (\ref{amxs}) as U(1) (approximate) pure gauge potential because
an electron has non-zero mass (even if it is very small) and hence it can not travel exactly at speed of light
$\beta^\mu_{ c}=(1,~{\vec \beta}_{ c}),~~{\vec \beta}^2_{c} =\frac{v^2}{c^2} = 1$ to produce the exact U(1) pure gauge
potential
\bea
A^\mu(x)=e\frac{\beta^\mu_{ c}}{\beta_{c} \cdot (x-X(\tau_0))}=\partial^\mu \omega(x),~~~~~~~~~\omega(x) = e~{\rm ln}[\beta_{ c} \cdot (x-X(\tau_0))].
\label{dmxk1}
\eea
From eq. (\ref{amxs}) one finds that the $\omega(x)$ appearing in the abelian pure gauge potential
\bea
A^\mu(x) = \partial^\mu \omega(x)
\label{bmxs}
\eea
in Maxwell theory is linearly proportional to $e$, {\it i.e.},
\bea
\omega(x) \propto e.
\label{cmxs}
\eea
Similarly one finds that the $\omega^a(x)$ appearing in the abelian-like non-abelian pure gauge potential
\bea
{\cal A}^{\mu a}(x) = \partial^\mu \omega^a(x)
\label{ybmxs}
\eea
in Yang-Mills theory is linearly proportional to $g$, {\it i.e.},
\bea
\omega^a(x) \propto g
\label{ycmxs}
\eea
whereas the SU(3) pure gauge potential $A^{\mu a}(x)$ in
\bea
T^aA^{\mu a}(x) = \frac{1}{ig}[\partial^\mu U(x)]U^{-1}(x),~~~~~~~~~~~~~~~~U(x)=e^{igT^a\omega^a(x)}
\label{su3pg}
\eea
in SU(3) Yang-Mills theory contains infinite number of terms upto infinite powers of $g$ and/or infinite
powers of $\omega^a(x)$. Note that eq. (\ref{ybmxs}) is only the first term in the expansion of $A^{\mu a}(x)$
in eq. (\ref{su3pg}).

\section{Charge and Charge Density of a point particle Using Quantum Mechanics }

In quantum mechanics the electron is described by the wave function $\psi(x)$. The Lagrangian density is given by
\bea
{\cal L}= -\frac{1}{4} F_{\mu \nu}F^{\mu \nu} + {\bar \psi} [i\gamma^\mu \partial_\mu -m +e\gamma^\mu A_\mu(x) ]\psi
\label{le}
\eea
from which we obtain the Dirac equation of the electron
\bea
[i\gamma^\mu \partial_\mu -m +e\gamma^\mu A_\mu(x) ]\psi(x) =0.
\label{de}
\eea
From eq. (\ref{le}) we find that the current density of the electron is given by
\bea
j^\mu(x) = e {\bar \psi}(x) \gamma^\mu \psi(x)
\label{ade}
\eea
which satisfies the continuity equation
\bea
\partial_\mu j^\mu(x)= 0.
\label{cceq}
\eea
Note that the Dirac wave function $\psi(x)$ in eqs. (\ref{de}) and (\ref{ade}) is yet to be
quantized in the sense of second quantization (field quantization) \cite{muta}.

Note that in classical mechanics the charge density of a point charge may be described by
delta function distribution, see eq. (\ref{kcurrent}), which implies that the charge density
at the position ${\vec x} \neq {\vec X}(t)$ is zero where ${\vec X}(t)$ is the spatial
position of the electron and ${\vec x}$ is the spatial position at which the current density
is determined. However, in quantum mechanics the probability
density $\psi^\dagger(t,{\vec x}) \psi(t,{\vec x})$ of finding a particle is not defined at a fixed position ${\vec x} = {\vec X}(t)$
but rather it is defined in a certain (infinitesimal) volume element. Hence one finds that in quantum mechanics the charge density
$e\psi^\dagger(t,{\vec x}) \psi(t,{\vec x})$ is not zero when ${\vec x} \neq {\vec X}(t)$.

However, when integrated over the entire (physically) allowed volume $V=\int d^3{\vec x}$, the charge density in
quantum mechanics and the charge density in classical mechanics reproduce the same charge $e$ of the
electron, {\it i.e.}, for the normalized wave function
\bea
\int d^3{\vec x}~{\psi}^\dagger(x) \psi(x)=1
\label{hqw}
\eea
we find from eq. (\ref{ade})
\bea
\int d^3{\vec x}~j_0(t,{\vec x})~=~ \int d^3{\vec x}~e~{\psi}^\dagger(x) \psi(x)~=~e.
\label{gqw}
\eea
Also from classical mechanics we find from eq. (\ref{kcurrent})
\bea
\int d^3{\vec x}~j_0(t,{\vec x})=~\int~d^3{\vec x}~e~\delta^{(3)}({\vec x}-{\vec X}(t))~=~e.
\label{gqw1}
\eea
Eqs. (\ref{gqw}) and (\ref{gqw1}) implies that the classical mechanics and quantum mechanics give the same
value of the electric charge $e$ of the electron even if the charge density distributions in classical mechanics
and in quantum mechanics are different.

As we have seen above, since the charge of a point particle
can not depend on the space coordinate ${\vec x}$, eqs. (\ref{ade}), (\ref{cceq}), (\ref{hqw}) and
(\ref{gqw}) suggest that quantum mechanics may provide a framework to determine the general form
of the fundamental charge. For example, in electrodynamics eqs. (\ref{ade}), (\ref{cceq}), (\ref{hqw})
and (\ref{gqw}) suggest that the fundamental electric charge $e$ of the electron can not be time dependent
but it has to be constant.

Let us apply this procedure to Yang-Mills theory to find the general form of eight time dependent fundamental
color charges $q^a(t)$ of a quark where $a=1,2,...,8$ are color indices.

\section{ Color Charge of the Quark in Yang-Mills Theory is Time Dependent }

In Yang-Mills theory the lagrangian density of the quark in the presence of non-abelian Yang-Mills potential
$A^{\mu a}(x)$ is given by \cite{yang,muta}
\bea
{\cal L}=-\frac{1}{4}F_{\mu \nu}^a(x)F^{\mu \nu a}(x) + {\bar \psi}_i(x)[i\delta_{ij} \gamma^\mu \partial_\mu -m\delta_{ij} +gT^a_{ij}\gamma^\mu A_\mu^a(x) ]\psi_j(x)
\label{yml}
\eea
where
\bea
F_{\mu \nu}^a(x)=\partial_\mu A_\nu^a(x)-\partial_\nu A_\mu^a(x) +gf^{abc}A_\mu^b(x) A_\nu^c(x),
\label{fmna}
\eea
with $a=1,2,...,8$ and $i,j=1,2,3$ in SU(3).

From eq. (\ref{yml}) one finds that the Dirac equation of the quark in the presence of non-abelian Yang-Mills
potential $A^{\mu a}(x)$ is given by
\bea
[i\delta_{ij} \gamma^\mu \partial_\mu -m\delta_{ij} +gT^a_{ij}\gamma^\mu A_\mu^a(x) ]\psi_j(x) =0.
\label{dquark}
\eea
Similarly from eq. (\ref{yml}) one finds that the Yang-Mills color current density $j^{\mu a}(x)$ generated by the
color charges of the quark in Yang-Mills theory is given by \cite{yang,muta}
\bea
j^{\mu a}(x) = g {\bar \psi}_i(x) \gamma^\mu T^a_{ij}\psi_j(x)
\label{ycur}
\eea
which satisfies the equation
\bea
D_\mu[A] j^{\mu a}(x) =0
\label{cce}
\eea
where
\bea
D_\mu^{ab}[A]=\delta^{ab}\partial_\mu +gf^{acb}A_\mu^c(x).
\label{dab}
\eea

Since eq. (\ref{cce}) is not a continuity equation ($\partial_\mu j^{\mu a}(x) \neq 0$)  one finds from eqs. (\ref{cce})
and (\ref{ycur}) that the color charge of the quark in Yang-Mills theory is not constant. As we have discussed above,
since the charge of a point particle can be obtained from the zero ($\mu=0$) component of a corresponding current density
after integrating over the entire (physically) allowed volume $V=\int d^3{\vec x}$, the color charge $q^a(t)$ of the quark
is time dependent. Since the color current density $j^{\mu a}(x)$ of a quark in eq. (\ref{ycur}) has eight color components
$a=1,2,...,8$ we find that there are eight time dependent fundamental color charges of a quark.

We denote eight time dependent fundamental color charges of a quark in Yang-Mills theory in SU(3)
by $q^a(t)$ where $a=1,2,...,8$ are color indices. These time dependent fundamental color charges $q^a(t)$ of the quark are
independent of quark flavor, {\it i.e.}, a color charge $q^a(t)$ of the $u$ (up) quark is same
as that of $d$ (down), $S$ (strange), $c$ (charm), $B$ (bottom) or $t$ (top) quark.

\section{Relation Between Fundamental Coupling Constant and Fundamental Charge in Maxwell Theory and in Yang-Mills Theory }

Let us now ask a fundamental question in the nature. Since the strong force or the color force is produced from
the color charges in the nature "what is the relation between the fundamental strong coupling constant $\alpha_s$
and the fundamental color charges $q^a(t)$ in the classical Yang-Mills theory" ?

One expects such a question because we know that a fundamental coupling constant is the measure of
the strength of a fundamental force in the nature produced by the fundamental charges.

For example the fundamental electromagnetic force in the nature is produced from the fundamental
electric charge $\pm e$. The strength of this fundamental electromagnetic force in the
nature is characterized by the typical value of the fundamental electromagnetic coupling
constant (or fine structure constant)
\bea
\alpha =\frac{(\pm e)^2}{4 \pi} \sim \frac{1}{137}
\label{ale}
\eea
which is related to the fundamental electric charge $\pm e$ in Maxwell theory.

Similarly, the fundamental strong force or the color force inside a hadron is produced from the fundamental
color charges $q^a(t)$ in the nature. Since the strength of the fundamental strong force or the color force is
characterized by the value of the fundamental strong coupling constant $\alpha_s$ one anticipates that
there is a relation between the fundamental strong coupling constant $\alpha_s$ and the fundamental color
charges $q^a(t)$ in Yang-Mills theory where $a=1,2,...,8$.

It is useful to denote $N^2-1$ color charges $q^a(t)$ of a fermion in SU(N) Yang-Mills theory
as components of a single color charge vector ${\vec q}(t)$ in $N^2-1$ dimensional color space (group space).
For example, in SU(3) Yang-Mills theory the eight time dependent fundamental color charges $q^a(t)$ of a quark can be
denoted by a single vector ${\vec q}(t)$ in the eight dimensional color space (group space in SU(3)) where
$a=1,2,...,8$. We call ${\vec q}(t)$ as eight-vector in color space,
similar to three-vectors ${\vec x}(t)$ and ${\vec v}(t)$ in coordinate space. Note that although the three-vector
${\vec x}(t)$ in coordinate space is not rotationally invariant, its magnitude $|{\vec x}(t)|$ is rotationally invariant. Similarly,
one finds that even if the eight-vector ${\vec q}(t)$ in color space is not gauge invariant, its magnitude $|{\vec q}(t)|$ is gauge
invariant where
\bea
{\vec q}^2(t)=|{\vec q}(t)|^2 = \sum_{a=1}^{8} q^a(t)q^a(t).
\label{gi}
\eea

In Maxwell theory the electromagnetic coupling constant
(or fine structure constant) is related to the magnitude $e^2$
of the fundamental electric charge '$- e$' of the electron via the equation
\bea
\alpha=\frac{(- e)^2}{4\pi}.
\label{alm}
\eea
Similarly, since the strong coupling constant $\alpha_s$ is gauge invariant,
one finds by extending eq. (\ref{alm}) to Yang-Mills theory that the strong coupling constant $\alpha_s$
is related to the magnitude ${\vec q}^2(t)$ of the time dependent fundamental color charge '${\vec q}(t)$'
of the quark via the equation
\bea
\alpha_s = \frac{{\vec q}^2(t)}{4 \pi}.
\label{aly}
\eea
Note that even if the fundamental color charge ${\vec q}(t)$ of the quark is time dependent,
the gauge invariant ${\vec q}^2(t) =\sum_{a=1}^{8} q^a(t)q^a(t)$ can be time independent.

In SU(2) Yang-Mills theory the color charge vector ${\vec q}(t)$ has three components $q_i(t)$ where $i=1,2,3$.

Since the color charge vector ${\vec q}(t)$ has eight components in SU(3) Yang-Mills theory we find from eq. (\ref{aly})
\bea
\alpha_s = \frac{q_1^2(t)+q_2^2(t)+q_3^2(t)+q_4^2(t)+q_5^2(t)+q_6^2(t)+q_7^2(t)+q_8^2(t)}{4 \pi}
\label{bly}
\eea
in SU(3) Yang-Mills theory.

Similarly, since the color charge vector ${\vec q}(t)$ has three components in SU(2) Yang-Mills theory
we find from eq. (\ref{aly})
\bea
\alpha_s = \frac{q_1^2(t)+q_2^2(t)+q_3^2(t)}{4 \pi}
\label{cly}
\eea
in SU(2) Yang-Mills theory.

Since the universal coupling $g$ in Yang-Mills theory and the strong
coupling constant $\alpha_s$ are related by
\bea
\alpha_s = \frac{g^2}{4 \pi}
\label{dly}
\eea
we find from eq. (\ref{aly}) that
\bea
g^2= {\vec q}^2(t).
\label{ely}
\eea

In SU(3) Yang-Mills theory we find from eq. (\ref{ely})
\bea
g^2= q_1^2(t)+q_2^2(t)+q_3^2(t)+q_4^2(t)+q_5^2(t)+q_6^2(t)+q_7^2(t)+q_8^2(t)
\label{fly}
\eea
and in SU(2) Yang-Mills theory we find from eq. (\ref{ely})
\bea
g^2= q_1^2(t)+q_2^2(t)+q_3^2(t).
\label{gly}
\eea

\section{ Color Current Density of Quark, Color Charge of Quark and General Form of Yang-Mills Potential }

Extending eq. (\ref{kcurrent}) of classical electrodynamics to time dependent color charge $q^a(t)$ in
classical chromodynamics we write
\bea
&& {\cal J}_0^a(t,{\vec x}) = q^a(t)~\delta^{(3)}({\vec x}-{\vec X}(t)) \nonumber \\
&& {\vec {\cal J}}^a(t,{\vec x}) = q^a(t)~{\vec v}(t)~\delta^{(3)}({\vec x}-{\vec X}(t))
\label{kcurrentn}
\eea
which satisfies the equation
\bea
\partial_\mu {\cal J}^{\mu a}(t,{\vec x}) = \frac{dq^a(t)}{dt}~\delta^{(3)}({\vec x}-{\vec X}(t)).
\label{hmn}
\eea
In a covariant formulation we find from eq. (\ref{kcurrentn})
\bea
{\cal J}^{\mu a}(x)=\int d\tau ~q^a(\tau)~u^\mu(\tau)~\delta^{(4)}(x-X(\tau)).
\label{mx8}
\eea
Similarly in covariant formulation we find from eq. (\ref{hmn})
\bea
\partial_\mu {\cal J}^{\mu a}(x) = \int d\tau \frac{dq^a(\tau)}{d\tau}~\delta^{(4)}(x-X(\tau))
\label{mx9}
\eea
which is not zero at $x^\mu =X^\mu(\tau)$ which confirms that the color charge $q^a(\tau)$ is not
constant.

The solution of the inhomogeneous wave equation
\bea
\partial^\nu \partial_\nu {\cal A}^{\mu a}(x)={\cal J}^{\mu a}(x)
\label{mx2v}
\eea
is given by
\bea
{\cal A}^{\mu a}(x)=\int d^4x' D_r(x-x'){\cal J}^{\mu a}(x')
\label{mx3}
\eea
where $D_r(x-x')$ is the retarded Greens function \cite{jackson}. When ${\cal A}^{\mu a}(x)$
satisfies the Lorentz gauge condition
\bea
\partial_\mu {\cal A}^{\mu a}(x) =0,
\label{mxl}
\eea
then eq. (\ref{mx2v}) reduces to the inhomogeneous Maxwell-like equation
\bea
\partial_\nu {\cal F}^{\nu \mu a}(x)={\cal J}^{\mu a}(x)
\label{mx1}
\eea
where
\bea
{\cal F}^{\mu \nu a}(x) =\partial^\mu {\cal A}^{\nu a}(x) -\partial^\nu {\cal A}^{\mu a}(x).
\label{afmn}
\eea

Hence we find that if ${\cal A}^{\mu a}(x)$ obtained from eq. (\ref{mx2v}) satisfies
Lorentz gauge condition eq. (\ref{mxl}) then it satisfies Maxwell-like eq. (\ref{mx1})
where ${\cal F}^{\mu \nu a}(x)$ is given by eq. (\ref{afmn}).

Using eq. (\ref{mx8}) in (\ref{mx3}) we find that
\bea
{\cal A}^{\mu a}(x)=q^a(\tau_0)\frac{u^\mu(\tau_0)}{u(\tau_0) \cdot (x-X(\tau_0))}
\label{mx10}
\eea
where $\tau_0$ is determined from the solution of the retarded condition as given by eq. (\ref{mx11}).

From eqs. (\ref{mx10}) and (\ref{mx11}) we find
\bea
&&\partial_\nu {\cal A}^{\mu a}(x) = q^a(\tau_0)\frac{ (x-X(\tau_0))_\nu {\dot u}^\mu(\tau_0)}{[u(\tau_0) \cdot (x-X(\tau_0))]^2}-q^a(\tau_0)\frac{ u^\mu(\tau_0)}{[u(\tau_0) \cdot (x-X(\tau_0))]^2}\nonumber \\
&&~[\frac{[\dot{u}(\tau_0) \cdot (x-X(\tau_0))-c^2](x-X(\tau_0))_\nu }{u(\tau_0) \cdot (x-X(\tau_0))}+u_\nu(\tau_0)]+[\partial_\nu q^a(\tau_0)] \frac{ u^\mu(\tau_0)}{u(\tau_0) \cdot (x-X(\tau_0))} \nonumber \\
\label{mxb}
\eea
which gives
\bea
\partial^\nu \partial_\nu {\cal A}^{\mu a}(x) =0
\label{mxe}
\eea
where
\bea
{\dot u}^\mu(\tau_0) =\frac{du^\mu(\tau)}{d\tau}|_{\tau=\tau_0}.
\label{udot}
\eea
Note that a quark has non-zero mass (even if the mass of the light quark is very small) which implies that a quark
can not travel exactly at speed of light $v=c$. As discussed in section II [see the paragraph after eq. (\ref{aafmn})],
at a common time $t=\frac{x_0}{c}=\frac{X_0(\tau)}{c}$ since ${\vec x} \neq {\vec X}(\tau)$ one finds that eq. (\ref{mxe})
is expected because eq. (\ref{mx10}) is obtained from eq. (\ref{mx2v}) where ${\cal J}^{\mu a}(x)$ is given by eq. (\ref{mx8}).

From eq. (\ref{mxb}) we find
\bea
\partial_\mu {\cal A}^{\mu a}(x) = \frac{{\dot q}^a(\tau_0)}{u(\tau_0) \cdot (x-X(\tau_0))},~~~~~~~~~~~~~~~~~{\dot q}^a(\tau_0)=\frac{dq^a(\tau)}{d\tau}|_{\tau=\tau_0}.
\label{mxnl}
\eea
When the quark in uniform motion is at its highest speed (which is arbitrarily
close to speed of light $v \sim c$) we find from eq. (\ref{mx10}) that ${\cal A}^{\mu a}(x)$
is given by
\bea
{\cal A}^{\mu a}(x)=q^a(\tau_0)\frac{\beta^\mu_{\sim c}}{\beta_{\sim c} \cdot (x-X(\tau_0))},~~~~~~~~~~\beta^\mu_{\sim c}=(1,~{\vec \beta}_{\sim c}),~~~~~~~~~~~~{\vec \beta}^2_{\sim c} =\frac{v^2}{c^2} \sim 1.
\label{mxk}
\eea
From eq. (\ref{mxnl}) we find that at all the time-space points $x^\mu$
(except at the spatial position ${\vec x}$ transverse to the motion of the quark
at the time of closest approach) the ${\cal A}^{\mu a}(x)$ in eq. (\ref{mxk}) satisfies
(approximate) Lorentz gauge condition
\bea
\partial_\mu {\cal A}^{\mu a}(x) \sim 0,
\label{mxi}
\eea
see section 4.2 of \cite{arxiv} for details.
From eqs. (\ref{mxe}), (\ref{mxi}) and (\ref{afmn}) we find that  at all the time-space points $x^\mu$
(except at the spatial position ${\vec x}$ transverse to the motion of the quark
at the time of closest approach) the ${\cal A}^{\mu a}(x)$ in eq. (\ref{mxk}) satisfies the equation
\bea
\partial_\nu {\cal F}^{\nu \mu a}(x)\sim 0.
\label{mxka}
\eea
As discussed above at a common time $t=\frac{x_0}{c}=\frac{X_0(\tau)}{c}$ since ${\vec x} \neq {\vec X}(\tau)$,
we find that eq. (\ref{mxka}) (approximately) satisfies Maxwell-like equation (\ref{mx1})
where ${\cal J}^{\mu a}(x)$ is given by eq. (\ref{mx8}).

From eqs. (\ref{mx10}) and (\ref{mx11}) we find
\bea
&&\partial^\mu {\cal A}^{\nu a}(x)-\partial^\nu {\cal A}^{\mu a}(x) = q^a(\tau_0)~\frac{(x-X(\tau_0))^\mu {\dot u}^\nu(\tau_0)- (x-X(\tau_0))^\nu {\dot u}^\mu(\tau_0)}{[u(\tau_0) \cdot (x-X(\tau_0))]^2}\nonumber \\
&&-q^a(\tau_0)[\frac{[\dot{u}(\tau_0) \cdot (x-X(\tau_0))-c^2][(x-X(\tau_0))^\mu u^\nu(\tau_0)-(x-X(\tau_0))^\nu u^\mu(\tau_0)]}{[u(\tau_0) \cdot (x-X(\tau_0))]^3}]\nonumber \\
&&+{\dot q}^a(\tau_0)\frac{ (x-X(\tau_0))^\mu u^\nu(\tau_0)-(x-X(\tau_0))^\nu u^\mu(\tau_0)}{[u(\tau_0) \cdot (x-X(\tau_0))]^2}
\label{mxoa}
\eea
which gives for uniform motion
\bea
&&\partial^\mu {\cal A}^{\nu a}(x)-\partial^\nu {\cal A}^{\mu a}(x) = c^2q^a(\tau_0)\frac{(x-X(\tau_0))^\mu u^\nu-(x-X(\tau_0))^\nu u^\mu}{[u \cdot (x-X(\tau_0))]^3}\nonumber \\
&&+{\dot q}^a(\tau_0)\frac{ (x-X(\tau_0))^\mu u^\nu-(x-X(\tau_0))^\nu u^\mu}{[u \cdot (x-X(\tau_0))]^2}
\label{mxo}
\eea
where ${\dot q}^a(\tau_0)$ is given by eq. (\ref{mxnl}).
When the quark in uniform motion is at its highest speed (which is arbitrarily close to the
speed of light $v \sim c$) we find from eq. (\ref{mxo}) that at all the time-space points $x^\mu$
(except at the spatial position ${\vec x}$ transverse to the motion of the quark
at the time of closest approach) the ${\cal A}^{\mu a}(x)$ in eq. (\ref{mxk})
gives
\bea
{\cal F}^{\mu \nu a}(x)\sim 0,
\label{mxq}
\eea
where ${\cal F}^{\mu \nu a}(x)$ is given by eq. (\ref{afmn}), see section 4.2 of \cite{arxiv} for details.

From eqs. (\ref{mxk}), (\ref{mxi}), (\ref{mxka}) and (\ref{mxq}) we find that at all the time-space points $x^\mu$
(except at the spatial position ${\vec x}$ transverse to the motion of the quark
at the time of closest approach) the
\bea
{\cal A}^{\mu a}(x)=q^a(\tau_0)\frac{\beta^\mu_{\sim c}}{\beta_{\sim c} \cdot (x-X(\tau_0))}
\label{mxr}
\eea
satisfies
\bea
\partial_\mu {\cal A}^{\mu a}(x) \sim 0,
\label{mxt}
\eea
\bea
\partial_\mu {\cal F}^{\mu \nu a}(x)\sim 0
\label{mxu}
\eea
and
\bea
{\cal F}^{\mu \nu a}(x)\sim 0
\label{mxv}
\eea
where ${\cal F}^{\mu \nu a}(x)$ is given by eq. (\ref{afmn}).

From eqs. (\ref{mxr}), (\ref{mxv}) and (\ref{afmn}) we find that at all the time-space points $x^\mu$
(except at the spatial position ${\vec x}$ transverse to the motion of the quark
at the time of closest approach) the eq. (\ref{mxr}) can be written as
\bea
{\cal A}^{\mu a}=q^a(\tau_0)\frac{\beta^\mu_{\sim c}}{\beta_{\sim c} \cdot (x-X(\tau_0))} \sim \partial^\mu \omega^a(x),~~~~~~~~\omega^a(x) = \int dl_c~ \frac{q^a(\tau_0)}{l_c},~~~~~~~~
l_c=\beta_{\sim c} \cdot (x-X(\tau_0))\nonumber \\
\label{mxs}
\eea
where $\int dl_c$ is an indefinite integration. Hence we find that the general expression of the
abelian-like non-abelian (approximate) pure gauge color potential at all the time-space points $x^\mu$
(except at the spatial position ${\vec x}$ transverse to the motion of the quark
at the time of closest approach) produced by the quark in uniform motion at its highest speed
(which is arbitrarily close to speed of light $v\sim c$) with time
dependent color charge $q^a(\tau)$ is given by eq. (\ref{mxs}).
Note that the expression of the abelian (approximate) pure gauge potential
at all the time-space points $x^\mu$
(except at the spatial position ${\vec x}$ transverse to the motion of the electron
at the time of closest approach) produced by the electron in uniform motion at its highest speed
(which is arbitrarily close to speed of light $v\sim c$) with constant
electric charge $e$ is given by eq. (\ref{amxs}).

We call eq. (\ref{mxs}) as abelian-like non-abelian (approximate) pure gauge color potential because
a quark has non-zero mass (even if the mass of the light quark is very small) and hence it can not travel exactly at speed of
light $\beta^\mu_c = (1,~{\vec \beta}_c)$ or ${\vec \beta}^2_c=\frac{v^2}{c^2} = 1$ to produce the
exact abelian-like non-abelian pure gauge color potential
\bea
{\cal A}^{\mu a}(x)=q^a(\tau_0)\frac{\beta^\mu_{ c}}{\beta_{c} \cdot (x-X(\tau_0))}=\partial^\mu \omega^a(x),~~~~~~~\omega^a(x) = \int dl_c~ \frac{q^a(\tau_0)}{l_c},~~~~~~~~
l_c=\beta_{c} \cdot (x-X(\tau_0)). \nonumber \\
\label{mxk1}
\eea

Note that in Maxwell theory the abelian pure gauge potential produced by the electron is given by
\bea
A^\mu(x) = \frac{1}{ie}[\partial^\mu U(x)]U^{-1}(x)=\partial^\mu \omega(x),~~~~~~~~~~~~~~~~~~U(x) = e^{ie\omega(x)}
\label{abpg}
\eea
where (see eq. (\ref{cmxs}))
\bea
\omega(x) \propto e.
\label{omgab}
\eea
In Yang-Mills theory the SU(3) non-abelian pure gauge potential $A^{\mu a}(x)$ in
\bea
T^aA^{\mu a}(x) = \frac{1}{ig}[\partial^\mu U(x)]U^{-1}(x),~~~~~~~~~~~~~~~~~U(x) = e^{igT^a\omega^a(x)}
\label{nab}
\eea
contains infinite number of terms upto infinite powers of $g$ and/or infinite powers of $\omega^a(x)$ where
\bea
\omega^a(x) \propto g.
\label{omgg}
\eea
The first term in the expansion in eq. (\ref{nab}) is the abelian-like non-abelian pure gauge color potential given by
\bea
{\cal A}^{\mu a}(x) = \partial^\mu \omega^a(x)
\label{nabpg}
\eea
where
\bea
\omega^a(x) \propto q^a(\tau_0) \propto g,
\label{qtg}
\eea
see eqs. (\ref{mxs}) and (\ref{ely}).

From eqs. (\ref{mxs}) and (\ref{nab}) we find that the general expression of the
SU(3) (approximate) pure gauge potential at all the time-space points $x^\mu$
(except at the spatial position ${\vec x}$ transverse to the motion of the quark
at the time of closest approach) produced by the the quark in uniform motion at its
highest speed (which is arbitrarily close to the speed of light $v\sim c$) in Yang-Mills theory is given by
\cite{arxiv}
\bea
A^{\mu a}(x) =  \frac{\beta^\mu_{\sim c}}{\beta_{\sim c} \cdot (x-X(\tau_0))}q^b(\tau_0) [\frac{e^{g\int dl_c \frac{Q(\tau_0)}{l_c}}-1}{g\int dl_c \frac{Q(\tau_0)}{l_c}}]_{ab}~ \sim [\partial^\mu \omega^b(x)]~ [\frac{e^{gM(x)}-1}{gM(x)}]_{ab}
\label{pfnab1}
\eea
where $\int dl_c$ is an indefinite integration and
\bea
Q^{ab}(\tau_0) =f^{abd}q^d(\tau_0),~~~~l_c= \beta_{\sim c} \cdot (x-X(\tau_0)),~~~~M_{ab}(x)=f^{abd}\omega^d(x),~~~~\omega^a(x) = \int dl_c~ \frac{q^a(\tau_0)}{l_c}.\nonumber \\
\label{prrr}
\eea
By making analogy with Maxwell theory we find from eq. (\ref{pfnab1}) that [see section 4.3 of \cite{arxiv} for details]
the general expression of the Yang-Mills potential (color potential) at all the time-space points $x^\mu$
produced by the color charges $q^a(\tau)$ of the quark in uniform motion at its highest speed (which is arbitrarily close
to the speed of light $v\sim c$) in Yang-Mills theory is given by
\bea
A^{\mu a}(x) =  \frac{\beta^\mu_{\sim c}}{\beta_{\sim c} \cdot (x-X(\tau_0))}q^b(\tau_0) [\frac{e^{g\int dl_c \frac{Q(\tau_0)}{l_c}}-1}{g\int dl_c \frac{Q(\tau_0)}{l_c}}]_{ab},~~~~Q^{ab}(\tau_0) =f^{abd}q^d(\tau_0),~~l_c= \beta_{\sim c} \cdot (x-X(\tau_0)).\nonumber \\
\label{pfnab}
\eea

From eq. (\ref{pfnab}) we find that [see section 5 of \cite{arxiv} for details]
the general form of the Yang-Mills potential (color potential) $A^{\mu a}(x)$
produced by the color charges $q^a(\tau)$ of the quark moving with arbitrary
four-velocity $u^\mu(\tau)=\frac{dX^\mu(\tau)}{d\tau}$ is given by \cite{arxiv}
\bea
A^{\mu a}(x) =  \frac{u^\mu(\tau_0)}{u(\tau_0) \cdot (x-X(\tau_0))}q^b(\tau_0) [\frac{e^{g\int dl \frac{Q(\tau_0)}{l}}-1}{g\int dl \frac{Q(\tau_0)}{l}}]_{ab}
\label{fnab}
\eea
where $\int dl$ is an indefinite integration,
\bea
Q^{ab}(\tau_0) =f^{abd}q^d(\tau_0),~~~~~~~~~~l= u(\tau_0) \cdot (x-X(\tau_0)),
\label{rrr}
\eea
$\tau_0$ is obtained from the solution of the retarded equation given by eq. (\ref{mx11}) and
the repeated color indices $b,d=1,2,...,8$ are summed.

From eqs. (\ref{fnab}), (\ref{yml}), (\ref{ycur}), (\ref{dab}) and (\ref{fmna}) we find that the non-abelian
Yang-Mills color current density $j^{\mu a}(x)$ of the quark which satisfies the equation
\bea
&& j^{\mu a}(x) =D_\nu[A]F^{\nu \mu a}(x) = \nonumber \\
&& \partial_\nu [ \partial^\nu A^{\mu a}(x) - \partial^\mu A^{\nu a}(x) +gf^{abc}A^{\nu b}(x)A^{\mu c}(x)]+gf^{abc}A_\nu^b
[\partial^\nu A^{\mu c}(x) - \partial^\mu A^{\nu c}(x) +gf^{chd}A^{\nu h}(x)A^{\mu d}(x)] \nonumber \\
\label{dcn}
\eea
contains infinite powers of $g$ or infinite powers of $q^a(\tau)$. Note that we have used the curly notations
${\cal A}^{\mu a}(x)$, ${\cal J}^{\mu a}(x)$ when ${\cal A}^{\mu a}(x)$, ${\cal J}^{\mu a}(x)$ are linearly
proportional to $q^a(\tau)$ [see eqs. (\ref{mxs}) and (\ref{mx8})] but we have used the usual notations
$A^{\mu a}(x)$, $j^{\mu a}(x)$ for the (full) non-abelian Yang-Mills theory where $A^{\mu a}(x)$, $j^{\mu a}(x)$
contain infinite powers of $q^a(\tau)$ [see eqs. (\ref{fnab}) and (\ref{dcn})]. Since the ${\cal J}^{\mu a}(x)$
in eq. (\ref{kcurrentn}) is linearly proportional to $g$ (see eq. (\ref{qtg})) and the non-abelian
Yang-Mills color current density $j^{\mu a}(x)$ of the quark in eq. (\ref{dcn}) contains infinite powers of $g$,
we find that $\int d^3{\vec x}~{\cal J}_0^a(t,{\vec x})$ in eq. (\ref{kcurrentn}) is different from
$\int d^3{\vec x}~j_0^a(t,{\vec x})$ in eq. (\ref{dcn}) in Yang-Mills theory. Note that they are same
for electron in Maxwell theory because $A^\mu(x)$ and $F^{\mu \nu}(x)$ are linearly proportional to the
electric charge $e$ of the electron in Maxwell theory.

From eqs. (\ref{ycur}), (\ref{dcn}) and (\ref{fnab}) we find that the non-abelian Yang-Mills color current density
\bea
j^{\mu a}(x)=g {\bar \psi}_i(x) \gamma^\mu T^a_{ij}\psi_j(x)
\label{aqwn}
\eea
of the quark in Yang-Mills theory contains infinite powers of $g$. Note that in Maxwell theory
the electromagnetic current density $j^\mu(x)=e {\bar \psi}(x) \gamma^\mu \psi(x)$ of the electron
is linearly proportional to $e$.

Hence we find that, unlike Maxwell theory where $\int d^3{\vec x}~{\psi}^\dagger(t,{\vec x}) \psi(t,{\vec x})=1$ is
independent of $e$, we find from eqs. (\ref{dcn}) and (\ref{aqwn}) that in Yang-Mills theory
$\int d^3{\vec x}~{\psi}^\dagger_i(t,{\vec x})T^a_{ij} \psi_j(t,{\vec x})$ contains infinite powers of $g$ even if
the wave function of the quark is normalized, {\it i.e.}, even if
\bea
\sum_{i=1}^3 \int d^3{\vec x}~{\psi}^\dagger_i(t,{\vec x})\psi_i(t,{\vec x})=1.
\eea

\section{ General Form of the Yang-Mills Potential (Color Potential) Produced by the Quark at Rest}

From the expression of the proper time
\bea
\tau = \int \frac{dX_0(\tau)}{c\gamma(X_0(\tau))}
\eea
we find that when ${\vec \beta}=\frac{{\vec v}}{c}=0$
\bea
\tau_0=\frac{X_0(\tau_0)}{c}.
\label{q1}
\eea
From the retarded condition from eq. (\ref{mx11}) we find
\bea
x_0-X_0(\tau_0)=ct-X_0(\tau_0)=r
\label{q2}
\eea
where
\bea
r=|{\vec x} -{\vec X}(\tau_0)|.
\label{ojmb}
\eea
From eqs. (\ref{q1}) and (\ref{q2}) we find
\bea
\tau_0 = t-\frac{r}{c}.
\label{q3}
\eea
Using eq. (\ref{q3}) in (\ref{fnab}) we find that the general form of the Yang-Mills potential (color potential) produced by
the color charges of the quark at rest is given by \cite{arxiv}
\bea
\Phi^a(x)=A_0^a(t,{\vec x}) =\frac{q^b(t-\frac{r}{c})}{r}\left[\frac{{\rm exp}[g\int dr \frac{Q(t-\frac{r}{c})}{r}]
-1}{g \int dr \frac{Q(t-\frac{r}{c})}{r}}\right]_{ab}
\label{upg1fg}
\eea
where $dr$ integration is an indefinite integration,
\bea
r=|{\vec x}-{\vec X}(\tau_0)|,~~~~~~~~Q_{ab}(\tau_0)=f^{abd}q^d(\tau_0),~~~~~~~~~~~\tau_0 = t-\frac{r}{c},
\label{uqg1}
\eea
${\vec X}(\tau_0)$ is the spatial position of the quark at rest at the retarded time
and the repeated color indices $b,d$(=1,2,...,8) are summed.

We find from eq. (\ref{upg1fg}) that, unlike Coulomb potential $A_0(t,{\vec x})=\frac{e}{|{\vec x}-{\vec X}|}$
produced by the electric charge of the electron at rest in Maxwell theory which is independent of time $t$, the color
potential $A_0^a(t,{\vec x})$ produced by the color charges of the quark in Yang-Mills theory depends on the retarded time
$\tau_0=t-\frac{r}{c}$ even if the quark is at rest \cite{arxiv}. This is a consequence of time dependent color charges
of the quark in Yang-Mills theory. The color potential (Yang-Mills potential) $A_0^a(t,{\vec x})$ at time $t$ produced by the
quark at rest depends on color charges $q^a(\tau_0)$ of the quark at the retarded time $\tau_0=t-\frac{r}{c}$.

In other words the color potential (Yang-Mills potential) $A_0^a(t,{\vec x})$
produced by the color charges $q^a(\tau_0)$ of the quark at rest from the spatial position ${\vec X}(\tau_0)$ not only
depends on the distance $r=|{\vec x}-{\vec X}(\tau_0)|$ but also depends on the retarded time
$\tau_0=t-\frac{r}{c}=t-\frac{|{\vec x}-{\vec X}(\tau_0)|}{c}$, see eq. (\ref{upg1fg}).

Note that when the color charge $q^a$ is constant we find from eq. (\ref{upg1fg})
\bea
\Phi^a(t, {\vec x}) =\frac{q^a}{|{\vec x}-{\vec X}|}
\label{ffg}
\eea
which reproduces the Coulomb-like potential (abelian-like potential), similar to Maxwell theory
where the constant electric charge $e$ produces Coulomb potential $\Phi(t, {\vec x})=A_0(t, {\vec x})=\frac{e}{|{\vec x}-{\vec X}|}$.
Hence from eq. (\ref{upg1fg}) we find that the the general form of the Yang-Mills potential (color potential) $A_0^a(t,{\vec x})$
produced by the color charges of the quark at rest is not abelian-like potential.

Hence one finds that the static systems in the Yang-Mills theory are, in general, not abelian-like.

\section{Yang-Mills Color Current Density of the Quark at Rest}

From eq. (\ref{fnab}) we find that the vector component of the Yang-Mills potential
(color potential) ${\vec A}^a(t, {\vec x})=0$ when the quark is at rest. However, since
the zero ($\mu =0$) component of the Yang-Mills potential (color potential) $A_0^a(t,{\vec x})$ in eq. (\ref{upg1fg})
produced by the quark at rest depends on the retarded time $\tau_0=t-\frac{r}{c}$, the Yang-Mills theory has some additional
features which are absent in Maxwell theory. For the quark at rest in Yang-Mills theory we find from eq. (\ref{dcn}) that
the zero ($\mu=0$) component of the Yang-Mills color current density $j_0^a(x)$ of the quark obeys the equation
\bea
\partial_0j_0^a(x)=\partial_0 \left[D_\mu[A]F^{\mu 0a}(x)\right]=\partial_0 \partial_i \partial^i A_0^a(t,{\vec x})\neq 0,~~~~~~~~~~i=1,2,3
\eea
because $A_0^a(t,{\vec x})$ in eq. (\ref{upg1fg}) depends on the retarded time $\tau_0=t-\frac{r}{c}$
even if the quark is at rest, {\it i.e.},
\bea
\partial_0j_0^a(x)\neq 0, ~~~~~~~~~{\rm for}~~{\rm quark}~~{\rm at}~~{\rm rest}.
\label{q0c}
\eea
On the other hand for electron at rest in Maxwell theory we find that the zero ($\mu=0$) component of the electromagnetic
current density $j_0(x)$ of the electron obeys the equation [see eq. (\ref{kcurrent1})]
\bea
\partial_0j_0(x)= 0,~~~~~~~~~{\rm for}~~{\rm electron}~~{\rm at}~~{\rm rest},
\eea
which is consistent with the fact that the electric charge $e$ of the
electron is constant. The eq. (\ref{q0c}) is consistent with the fact that in Yang-Mills theory the color
charge $q^a(\tau)$ of the quark is time dependent and hence the
color potential (Yang-Mills potential) $A_0^a(t,{\vec x})$ at time $t$ produced by the
quark at rest depends on the retarded time $\tau_0=t-\frac{r}{c}$, see eq. (\ref{upg1fg}).

For the quark at rest in Yang-Mills theory we find from eq. (\ref{dcn}) that the vector component of
the Yang-Mills color current density ${\vec j}^a(x)$ of the quark is given by
\bea
j^{ia}(x)=D_\mu[A]F^{\mu ia}(x)=-\partial_0 \partial^i A_0^a(t,{\vec x})-gf^{abc}A_0^b(t,{\vec x})\partial^i A_0^c(t,{\vec x})\neq 0,~~~~~~~~~~i=1,2,3\nonumber \\
\eea
even if the quark is at rest, {\it i.e.},
\bea
{\vec j}^a(x)\neq 0,~~~~~~~~~{\rm for}~~{\rm quark}~~{\rm at}~~{\rm rest},
\eea
even if ${\vec {\cal J}}^a(x)=0$ from eq. (\ref{kcurrentn}). On the other hand for electron at rest in Maxwell theory we find
that the vector component of the electromagnetic current density ${\vec j}(x)$ of the electron is given by
[see eq. (\ref{kcurrent})]
\bea
{\vec j}(x)= 0,~~~~~~~~~{\rm for}~~{\rm electron}~~{\rm at}~~{\rm rest}.
\eea
Hence one finds that the static systems in the Yang-Mills theory are, in general, not abelian-like.

\section{ Form of Fundamental Charge From Dirac Wave Function in Maxwell Theory and in Yang-Mills Theory }

Note that since the color current density
$j^{\mu a}(x)=g{\bar \psi}_i(x) \gamma^\mu T^a_{ij}\psi_j(x)$ of the quark  in
Yang-Mills theory is not gauge invariant, the color charge $q^a(t)$ of the quark is not gauge
invariant and is not a physically measurable quantity but its magnitude
${\vec q}^2(t)=\sum_{a=1}^8q^a(t)q^a(t)=g^2$ is gauge invariant and is a physically measurable quantity.
As mentioned in section V, the time dependent color charge vector ${\vec q}(t)$ of a fermion in SU(N) Yang-Mills
theory is a $N^2-1$ dimensional (real) vector. The time dependent gauge transformation of the color
charge $q^a(t)$ of the quark in the adjoint representation of SU(3) is given by
\begin{equation}
q'^a(t)=R_{ab}(t)q^b(t),~~~~~~~~~~~R_{ab}(t)=[e^{S(t)}]_{ab},~~~~~~~~~~~~~S_{ab}(t)=f^{abc}\beta^c(t)
\label{r3}
\end{equation}
where $\beta^a(t)$ are time dependent parameters,
$f^{abc}$ are the antisymmetric structure constants of the SU(3) group
and $a,b,c=1,2,...,8$. Similarly, the time dependent gauge transformation of the color
charge $q_i(t)$ of a fermion in the adjoint representation of SU(2) is given by
\begin{equation}
q'_i(t)=R_{ij}(t)q_j(t),~~~~~~~~~~~R_{ij}(t)=[e^{S(t)}]_{ij},~~~~~~~~~~~~~S_{ij}(t)=\epsilon_{ijk}\beta_k(t)
\label{r2}
\end{equation}
where $\epsilon_{ijk}$ are the antisymmetric structure constants of the SU(2) group
and $i,j,k=1,2,3$. One may recall that
the three generators of the SO(3) group are proportional to the three $3\otimes 3$ matrices
$G^i_{jk}=\epsilon_{ijk}$ respectively, which implies that the gauge transformation in
the adjoint representation of SU(2) corresponds to a rotation in SO(3), see for example
\cite{creutz}. Hence we find that the time dependent gauge transformation of the color charge
(real) three-vector ${\vec q}(t)$ of a fermion in
the adjoint representation of SU(2) in eq. (\ref{r2})
corresponds to a time dependent rotation in SO(3). The gauge transformation in
SU(2) may be described in terms of three Euler angles by reparameterising the
gauge transformation, see for example \cite{euler}. Similarly the gauge transformation in
SU(3) may be described in terms of eight Euler angles by reparameterising the gauge
transformation. The geometry of SU(3)
is described in terms of 8 Euler angles in \cite{byrd} and the 64 components of the $8\otimes 8$ arbitrary matrix
in the adjoint representation of SU(3) is obtained  in terms of 8 Euler angles in \cite{byrd2}.
It is useful to remember that, although the gauge transformation in
the adjoint representation of SU(2) corresponds to a rotation in SO(3), the gauge transformation in
the adjoint representation of SU(3) does not correspond to a general rotation in SO(8)
because a general rotation in SO(8) is not described by 8 real parameters but
a general rotation in SO(8) is described by 28 real parameters, see for example \cite{yeh}.
Hence one finds that the time dependent gauge transformation in
the adjoint representation of SU(3) in eq. (\ref{r3}) does not
correspond to a time dependent general rotation in SO(8) although the time dependent gauge
transformation in the adjoint representation of SU(2) in eq. (\ref{r2}) corresponds to a time dependent rotation in SO(3).
Hence one expects that in the Yang-Mills theory the general form of three time dependent color
charges $q_i(t)$ of a fermion in SU(2) and the general form of eight time dependent fundamental color charges
$q^a(t)$ of the quark in SU(3) may reflect the fact that SU(2) and SO(3) are locally isomorphic, while
SU(3) and SO(8) are not [see section XV for more discussion on this]. Since color charge $q^a(t)$ of the quark is not
gauge invariant it does not mean that we can not obtain the form of the color charge $q^a(t)$ of the quark in Yang-Mills
theory. For example, we know that the Maxwell potential $A^\mu(x)$ (the Lienard-Weichert potential) is not gauge invariant (only
$F^{\mu \nu}(x)$ is gauge invariant, see eq. (\ref{aafmn})), but
that does not mean that we can not derive the form of the Maxwell potential (Lienard-Weichert potential)
in Maxwell theory by solving Maxwell equation using Green's function technique, see \cite{jackson}.
Similarly, one finds that one can obtain the form of the color charge
$q^a(t)$ of the quark in Yang-Mills theory even if the color charge $q^a(t)$ of the quark is not gauge invariant.
Any gauge invariant quantities which are related to physically measurable quantities remain unchanged under the
gauge transformation of the fundamental color charge $q^a(t)$ of the quark.

In Maxwell theory the electric current density $j^\mu(x)$ of the electron and the fundamental electric
charge $e$ of the electron are gauge invariants. The form of the fundamental electric charge $e$ of the
electron can be obtained from the zero component $j_0(x)$ of the electric current density $j^\mu(x)$ of
the electron by using the Dirac wave function $\psi(x)$ of the electron (see below).
In Yang-Mills theory the color current density $j^{\mu a}(x)$ of the quark
and the fundamental color charge $q^a(t)$ of the quark are not gauge invariants.
Hence in analogy to Maxwell theory one can obtain the form of gauge non-invariant fundamental color charge $q^a(t)$
of the quark from the zero component $j_0^a(x)$ of the gauge non-invariant color current density $j^{\mu a}(x)$
of the quark by using Dirac wave function $\psi_i(x)$ of the quark (see below). As mentioned in section III
these Dirac wave functions $\psi(x)$ of the electron and $\psi_i(x)$ of the quark are yet to be quantized in
the sense of second quantization (field quantization), see \cite{muta}.

Note that in Maxwell theory even if scalar ($\mu=0$) component
\begin{equation}
j_0(x) = e {\bar \psi}(x) \gamma_0 \psi(x)=e\psi^\dagger(x) \psi(x)
\label{scd}
\end{equation}
of the electric current density $j^\mu(x)=e{\bar \psi}(x)\gamma^\mu \psi(x)$ of the electron
is proportional to $\psi^\dagger(x) \psi(x)$, the number
density of the electron, it does not naturally give constant electric charge $e$ of the electron after
integrating over full spatially volume available for the electron unless we implement the continuity equation
as given by eq. (\ref{2c}). This is because if the continuity equation is not satisfied, {\it i.e.}, if
\begin{equation}
\partial_\mu j^\mu(x) \neq 0
\label{3c}
\end{equation}
then the electric charge $e$ of the electron becomes time dependent instead of being constant.

Hence in order to determine the form of the fundamental color charge $q^a(t)$ of the quark from the
Dirac wave function $\psi_i(x)$ of the quark in Yang-Mills theory we make analogy with the Maxwell theory.
First of all we notice that the Yang-Mills theory was developed in analogy to the procedure in electromagnetic
theory (Maxwell theory) by extending the U(1) gauge group to the SU(3) gauge group appropriately,
see \cite{yang}. For example in analogy to $D_\mu \psi = (\partial_\mu -ie A_\mu)\psi$ in
electromagnetic theory (Maxwell theory) all derivatives of $\psi$ in Yang-Mills theory appear
in the combination $D_\mu \psi = (\partial_\mu -i\epsilon B_\mu)\psi$ where $B_\mu =T^a A_\mu^a$,
see \cite{yang}. Similarly, in analogy to commutator equation $[D_\mu,~D_\nu]=-ieF_{\mu \nu}$ in
electromagnetic theory (Maxwell theory) which predicts the relation between
$F_{\mu \nu}(x)$ and $A_\mu(x)$ as given by eq. (\ref{aafmn})
one finds that in Yang-Mills theory the commutator equation
$[D_\mu,~D_\nu]=-igT^aF_{\mu \nu}^a$ predicts the relation between $F_{\mu \nu}^a(x)$
and $A_\mu^a(x)$ as given by eq. (\ref{fmna}), see \cite{yang,muta}. Similarly, in analogy to the
electromagnetic field Lagrangian density ${\cal L}_{\rm EM}=-\frac{1}{4} F_{\mu \nu}F^{\mu \nu}$ in
electromagnetic theory (Maxwell theory) where $F_{\mu \nu}(x)$ is given by eq. (\ref{aafmn}) one writes down
the Yang-Mills field Lagrangian density ${\cal L}_{\rm YM}=-\frac{1}{4} F_{\mu \nu}^aF^{\mu \nu a}$ in
Yang-Mills theory where $F_{\mu \nu}^a(x)$ is given by eq. (\ref{fmna}), see \cite{yang}.
Similarly, in analogy to electromagnetic theory (Maxwell theory) where the form of the
physical electromagnetic potential $A^\mu(x)$ (Lienard-Wiechert potential or Coulomb potential) produced by the
electric charge $e$ of the electron can be obtained from the form of U(1) pure gauge potential produced
by the electron, one finds that the form of the color potential (Yang-Mills potential) $A^{\mu a}(x)$
produced by the color charges $q^a(\tau)$ of the quark in Yang-Mills theory can be obtained
from the form of SU(3) pure gauge potential produced by the quark, see \cite{arxiv}.
Similarly, in order to determine the form of the fundamental color charge $q^a(t)$ of the quark from
the Dirac wave function $\psi_i(x)$ of the quark we make
analogy with the Maxwell theory. First of all we notice from eqs. (\ref{scd}), (\ref{ade}) and (\ref{2c})
that in Maxwell theory the form of the fundamental electric charge $e$ of the electron can be determined
from the Dirac wave function $\psi(x)$ of the electron by using,
1) the continuity equation $\partial_\mu j^\mu(x) = 0$ of the electric current density $j^\mu(x)$ of the electron,
2) scalar ($\mu =0$) component of the electric current density $j^\mu(x) = e{\bar \psi}(x) \gamma^\mu \psi(x)$ of the electron and
3) number density $n_e(x)=\psi^\dagger(x) \psi(x)$ of the electron by using the normalization condition
$\int d^3x \psi^\dagger(x) \psi(x)=1$. Hence in analogy to Maxwell theory, one finds in Yang-Mills theory
that the form of the fundamental color charge $q^a(t)$ of the quark can be determined from the Dirac
wave function $\psi_i(x)$ of the quark by using, 1) the equation $D_\mu[A] j^{\mu a}(x) = 0$ of the
color current density $j^{\mu a}(x)$ of the quark, 2) scalar ($\mu=0$) component of the color current
density $j^{\mu a}(x) = g{\bar \psi}_i(x) \gamma^\mu T^a_{ij} \psi_j(x)$ of the quark
and 3) number density $n_q(x)=\psi^\dagger_i(x) \psi_i(x)$ of the quark by using the normalization condition
$\int d^3x \psi^\dagger_i(x) \psi_i(x)=1$ (see below).

\section{Fermion Color Current Density, Fermion Wave Function and Pauli Matrices in Yang-Mills Theory in SU(2) }

In SU(2) Yang-Mills theory the Yang-Mills color current density of a fermion is given by
\bea
j^{\mu i}(x) = g {\bar \psi}_k(x) \gamma^\mu \tau^i_{kn} \psi_n(x);~~~~~~~~~i=1,2,3;~~~~~~~~~~k,n=1,2;
\label{2sa}
\eea
where three generators $\tau^i$ of the SU(2) group are given by
\bea
\tau_1=\frac{1}{2}\left( \begin{array}{cc}
0&,1\\
1&,0
\end{array} \right),~~~\tau_2=\frac{1}{2}\left( \begin{array}{cc}
0&,-i\\
i&,0
\end{array} \right),~~~\tau_3=\frac{1}{2}\left( \begin{array}{cc}
1&,0\\
0&,-1
\end{array} \right)
\label{paulim}
\eea
which are related to Pauli matrices $\sigma^i$ via the relation $\tau^i=\frac{\sigma^i}{2}$.

If there is only one fermion in the entire (physically) allowed volume
$V=\int d^3{\vec x}$ then the total probability of finding that fermion
in the entire volume is 1. This implies that the normalized wave functions
$\psi_k(x)$ of the fermion, where $k=1,~2$ are the color indices of the fermion
wave function in SU(2) Yang-Mills theory, satisfy the normalization condition
\bea
\int d^3{\vec x}~[\psi_1^\dagger(x) \psi_1(x) + \psi_2^\dagger(x) \psi_2(x) ] =1.
\label{wn}
\eea
From eqs. (\ref{2sa}) and (\ref{paulim}) we find
\bea
&& \int d^3{\vec x} j_0^1(x) = \frac{g}{2}\int d^3{\vec x} \psi^\dagger_1(x) \psi_2(x)+\frac{g}{2}\int d^3{\vec x} \psi^\dagger_2(x) \psi_1(x), \nonumber \\
&& \int d^3{\vec x}j_0^2(x) = -\frac{ig}{2}\int d^3{\vec x} \psi^\dagger_1(x) \psi_2(x)+\frac{ig}{2}\int d^3{\vec x} \psi^\dagger_2(x) \psi_1(x),\nonumber \\
&& \int d^3{\vec x}j_0^3(x) = \frac{g}{2}\int d^3{\vec x} \psi^\dagger_1(x) \psi_1(x)-\frac{g}{2}\int d^3{\vec x} \psi^\dagger_2(x) \psi_2(x).
\label{2sd}
\eea

As we have seen above, since $\int d^3{\vec x}~{\psi}^\dagger_i(t,{\vec x})\tau^a_{ij} \psi_j(t,{\vec x})$ in
Yang-Mills theory is non-linear function of $g$ we write
\bea
d_{11}(t,g) = \int d^3{\vec x} \psi^\dagger_1(x) \psi_1(x),~~~~~~~~~d_{22}(t,g) = \int d^3{\vec x} \psi^\dagger_2(x) \psi_2(x),~~~~~~~~~d_{12}(t,g) = \int d^3{\vec x} \psi^\dagger_1(x) \psi_2(x)\nonumber \\
\label{dg}
\eea
where
$d_{11}(t,g)$, $d_{22}(t,g)$ are $t$ and $g$ dependent real positive functions and $d_{12}(t,g)$ is $t$
and $g$ dependent complex function.

From eqs. (\ref{2sd}) and (\ref{dg}) we find
\bea
&&\int d^3{\vec x} j_0^1(x)  = \frac{g}{2}\times [d_{12}(t,g)+d^{*}_{12}(t,g)] \nonumber \\
&&\int d^3{\vec x} j_0^2(x) = -\frac{ig}{2}\times [d_{12}(t,g)-d^{*}_{12}(t,g)] \nonumber \\
&& \int d^3{\vec x} j_0^3(x)= \frac{g}{2}\times [ d_{11}(t,g)-d_{22}(t,g)].
\label{a2shj}
\eea
From the normalization condition, see eq. (\ref{wn}), we find
\bea
d_{11}(t,g)+d_{22}(t,g)=1
\label{wn1}
\eea
where we have used eq. (\ref{dg}).

Since $d_{11}(t,g)$ and $d_{22}(t,g)$ are $t$ and $g$ dependent real positive functions (see eq. (\ref{dg}))
we can write eq. (\ref{wn1}) as
\bea
d_{11}(t,g) = {\rm cos}^2\Theta(t,g),~~~~~~~~~~~~~~~~~~d_{22}(t,g) = {\rm sin}^2\Theta(t,g)
\label{wn2}
\eea
where
\bea
0 \le \Theta(t,g) \le 2\pi.
\label{thtb}
\eea

Using eq. (\ref{wn2}) in (\ref{a2shj}) we find
\bea
&&\int d^3{\vec x} j_0^1(x)  = \frac{g}{2}\times [d_{12}(t,g)+d^{*}_{12}(t,g)] \nonumber \\
&&\int d^3{\vec x} j_0^2(x) = -\frac{ig}{2}\times [d_{12}(t,g)-d^{*}_{12}(t,g)] \nonumber \\
&& \int d^3{\vec x} j_0^3(x)= \frac{g}{2}\times {\rm cos}[2\Theta(t,g)].
\label{a2sh}
\eea

\section{ General Form of Fundamental Color Charge of a Fermion in Yang-Mills Theory in SU(2) }

Since the fundamental color charge vector ${\vec q}(t)$ is linearly proportional to $g$ (see eqs.
(\ref{ely}) and (\ref{qtg})) we find from eq. (\ref{a2sh}) that the color charge $q_i(t)$ of a fermion
in Yang-Mills theory in SU(2) takes the form
\bea
&& q_1(t) =  \frac{g}{2}\times [d_{12}(t)+d^{*}_{12}(t)] \nonumber \\
&& q_2(t) = -\frac{ig}{2}\times [d_{12}(t)-d^{*}_{12}(t)] \nonumber \\
&& q_3(t) = \frac{g}{2}\times {\rm cos}[2\Theta(t)]
\label{a2shna}
\eea
where the complex function $d_{12}(t)$ and the real phase factor $\Theta(t)$ depend on time
$t$ but are independent of $g$. Since $d_{12}(t)$ is a complex function we can write eq.
(\ref{a2shna}) as
\bea
&& q_1(t) =  g \times |d_{12}(t)|\times {\rm cos}\phi(t) \nonumber \\
&& q_2(t) = g \times |d_{12}(t)|\times {\rm sin}\phi(t) \nonumber \\
&& q_3(t) = \frac{g}{2}\times {\rm cos}[2\Theta(t)]
\label{a2shn}
\eea
where the principal argument
\bea
{\rm Arg}\left(d_{12}(t)\right)=\phi(t) = {\rm tan}^{-1}\left[\frac{{\rm Im}[d_{12}(t)]}{{\rm Re}[d_{12}(t)]}\right].
\label{phio}
\eea
of the complex function $d_{12}(t)$ lying in the range
\bea
-\pi < \phi(t) \le  \pi.
\label{phin}
\eea
It is important to remember that the real phases $\Theta(t)$ and $\phi(t)$ in eq. (\ref{a2shn})
are not independent of time $t$. This is because if the real phases $\Theta(t)$ and $\phi(t)$ are
independent of time $t$ then the non-abelian Yang-Mills potential $A^{\mu a}(x)$
in eq. (\ref{fnab}) becomes Maxwell-like potential $A^\mu(x)$.

From eqs. (\ref{gly}) and (\ref{a2shn}) we find
\bea
|d_{12}(t)|^2 + \frac{1}{4}{\rm cos}^2[2\Theta(t)] =1.
\label{b2shj}
\eea
From eq. (\ref{thtb}) we find that the maximum allowed range of $\Theta(t)$ is
\bea
0 \le \Theta(t) \le 2\pi.
\label{thtbm}
\eea
From eqs. (\ref{b2shj}) and (\ref{thtbm}) we find
\bea
\frac{\sqrt{3}}{2}\le |d_{12}(t)| \le 1.
\label{b2sh}
\eea
We write
\bea
\frac{1}{4}{\rm cos}^2[2\Theta(t)] = {\rm cos}^2\theta(t)
\label{tht1}
\eea
where from eqs. (\ref{thtbm}) and (\ref{tht1}) we find
\bea
0 \le {\rm cos}^2\theta(t) \le \frac{1}{4}.
\label{tht2}
\eea
Since $|d_{12}(t)|$ is positive we find from eqs. (\ref{b2shj}), (\ref{b2sh}), (\ref{tht1}) and (\ref{tht2})
that we can write
\bea
|d_{12}(t)|={\rm sin}\theta(t)
\label{snth}
\eea
where
\bea
\frac{\pi}{3} \le \theta(t) \le \frac{2\pi}{3}.
\label{tht3}
\eea
Hence from eqs. (\ref{tht1}), (\ref{snth}), (\ref{tht3}), (\ref{phin}) and (\ref{a2shn}) we find that
the general form of three time dependent fundamental color charges of a fermion in Yang-Mills theory
in SU(2) is given by
\bea
&& q_1(t) = g\times {\rm sin}\theta(t)\times {\rm cos}\phi(t), \nonumber \\
&& q_2(t) = g\times {\rm sin}\theta(t)\times {\rm sin}\phi(t), \nonumber \\
&& q_3(t) = g\times {\rm cos}\theta(t)
\label{2skfn}
\eea
which reproduces eq. (\ref{2skin}) where
\bea
\frac{\pi}{3} \le \theta(t) \le \frac{2\pi}{3},~~~~~~~~~~~~-\pi < \phi(t) \le  \pi.
\label{2fna}
\eea
Note that if these two phases become constants then
the three color charges $q_i$ of the fermion become constants in which case
the Yang-Mills potential $A^{\mu i}(x)$ reduces to Maxwell-like (abelian-like) potential
\cite{arxiv}. It should be remembered that the
static systems in the Yang-Mills theory are, in general, not
abelian-like. For example, unlike Maxwell theory where the electron at rest produces
(abelian) Coulomb potential, the fermion at rest in Yang-Mills theory does not
produce abelian-like potential \cite{arxiv}.

The time dependence must be retained in order
not to end up in a quasi-abelian setting. In view of the fact that the whole
derivation boils down to a gauge transformation of a given spatially integrated
charge configuration this is also no surprise. The only aspect that cannot be
captured in a quasi-abelian representation is the time dependence of the
non-abelian gauge transformation.

\section{Yang-Mills Color Current Density of Quark, the Wave Function of Quark and Gell-Mann Matrices in Yang-Mills Theory in SU(3) }

Eight generators $T^a$ of the SU(3) group in the Yang-Mills theory are given by
\bea
&&T_1 =\frac{1}{2} \left(\begin{array}{ccc}
0,&1,&0\\
1,&0,&0\\
0,&0,&0
\end{array} \right),~~~~
T_2 =\frac{1}{2} \left(\begin{array}{ccc}
0,&-i,&0\\
i,&0,&0\\
0,&0,&0
\end{array} \right),~~~~
T_3 =\frac{1}{2} \left(\begin{array}{ccc}
1,&0,&0\\
0,&-1,&0\\
0,&0,&0
\end{array} \right),~~~~
T_4 =\frac{1}{2} \left(\begin{array}{ccc}
0,&0,&1\\
0,&0,&0\\
1,&0,&0
\end{array} \right), \nonumber \\
&& T_5 =\frac{1}{2}\left(\begin{array}{ccc}
0,&0,&-i\\
0,&0,&0\\
i,&0,&0
\end{array} \right),~~~~
T_6 =\frac{1}{2} \left(\begin{array}{ccc}
0,&0,&0\\
0,&0,&1\\
0,&1,&0
\end{array} \right),~~~~
T_7 =\frac{1}{2} \left(\begin{array}{ccc}
0,&0,&0\\
0,&0,&-i\\
0,&i,&0
\end{array} \right),~~~~
T_8 =\frac{1}{2\sqrt{3}} \left(\begin{array}{ccc}
1,&0,&0\\
0,&1,&0\\
0,&0,&-2
\end{array} \right)\nonumber \\
\label{T}
\eea
which are related to Gell-Mann matrices $\lambda^a$ via the relation $T^a=\frac{\lambda^a}{2}$.
In Yang-Mills theory in SU(3) the Yang-Mills color current density of the quark is given by
\bea
j^{\mu a}(x) = g {\bar \psi}_i(x) \gamma^\mu T^a_{ij} \psi_j(x);~~~~~~~~~~~~a=1,2,...,8;~~~~~~~~i,j=1,2,3.
\label{q2sa}
\eea
The normalized wave functions $\psi_i(x)$ of the quark obey the equation
\bea
\int d^3{\vec x}~ \psi^\dagger_1(x)\psi_1(x) + \int d^3{\vec x}~ \psi^\dagger_2(x)\psi_2(x) +\int d^3{\vec x}~ \psi^\dagger_3(x)\psi_3(x) =1.
\label{nwq}
\eea
From eqs. (\ref{q2sa}) and (\ref{T}) we find
\bea
&& \int d^3{\vec x} j_0^1(x)= \frac{g}{2}[h_{12}(t,g)+h^{*}_{12}(t,g)], \nonumber \\
&& \int d^3{\vec x} j_0^2(x) = -\frac{ig}{2}[h_{12}(t,g)-h^{*}_{12}(t,g)],\nonumber \\
&& \int d^3{\vec x} j_0^3(x) = \frac{g}{2}[h_{11}(t,g)-h_{22}(t,g)], \nonumber \\
&& \int d^3{\vec x} j_0^4(x) = \frac{g}{2}[h_{13}(t,g)+h^{*}_{13}(t,g)],\nonumber \\
&& \int d^3{\vec x} j_0^5(x) = -\frac{ig}{2}[h_{13}(t,g)-h^{*}_{13}(t,g)],\nonumber \\
&& \int d^3{\vec x} j_0^6(x) = \frac{g}{2}[h_{23}(t,g)+h^{*}_{23}(t,g)], \nonumber \\
&& \int d^3{\vec x} j_0^7(x) = -\frac{ig}{2}[h_{23}(t,g)-h^{*}_{23}(t,g)],\nonumber \\
&& \int d^3{\vec x} j_0^8(x) = \frac{g}{2\sqrt{3}}[ h_{11}(t,g)+h_{22}(t,g) -2h_{33}(t,g)]
\label{2shas}
\eea
where
\bea
&& h_{11}(t,g) = \int d^3{\vec x} \psi^\dagger_1(x) \psi_1(x),~~~~~~~~~h_{22}(t,g) = \int d^3{\vec x} \psi^\dagger_2(x) \psi_2(x),~~~~~~~~~h_{33}(t,g) = \int d^3{\vec x} \psi^\dagger_3(x) \psi_3(x), \nonumber \\
&&h_{12}(t,g) = \int d^3{\vec x} \psi^\dagger_1(x) \psi_2(x),~~~~~~~~~h_{13}(t,g) = \int d^3{\vec x} \psi^\dagger_1(x) \psi_3(x),~~~~~~~~~h_{23}(t,g) = \int d^3{\vec x} \psi^\dagger_2(x) \psi_3(x) \nonumber \\
\label{qdg}
\eea
and from eq. (\ref{nwq}) we find
\bea
h_{11}(t,g)+h_{22}(t,g)+h_{33}(t,g)=1.
\label{qwn1}
\eea
From eq. (\ref{qdg}) we find that $h_{11}(t,g)$, $h_{22}(t,g)$, $h_{33}(t,g)$ are $t$ and $g$ dependent
real positive functions and $h_{12}(t,g)$, $h_{13}(t,g)$ and $h_{23}(t,g)$ are $t$ and $g$ dependent complex
functions.

From now onwards we can proceed exactly in the way similar to SU(2).

Since $h_{11}(t,g)$, $h_{22}(t,g)$ and $h_{33}(t,g)$ are $t$ and $g$ dependent real positive functions (see eq. (\ref{qdg}))
we can write eq. (\ref{qwn1}) as
\bea
h_{11}(t,g) ={\rm sin}^2\Theta(t,g)\times {\rm cos}^2\Phi(t,g),~~~~~h_{22}(t,g) ={\rm sin}^2\Theta(t,g)\times {\rm sin}^2\Phi(t,g),~~~~h_{33}(t,g) = {\rm cos}^2\Theta(t,g)\nonumber \\
\label{qwn2}
\eea
where
\bea
0 \le \Theta(t,g) \le \pi,~~~~~~~~~~~~~~~~0 \le \Phi(t,g) < 2\pi.
\label{qthtb}
\eea
Using eq. (\ref{qwn2}) in (\ref{2shas}) we find
\bea
&& \int d^3{\vec x} j_0^1(x)= \frac{g}{2}\times [h_{12}(t,g)+h^{*}_{12}(t,g)], \nonumber \\
&& \int d^3{\vec x} j_0^2(x) = -\frac{ig}{2}\times[h_{12}(t,g)-h^{*}_{12}(t,g)],\nonumber \\
&& \int d^3{\vec x} j_0^3(x) = \frac{g}{2}\times{\rm sin}^2\Theta(t,g) \times{\rm cos}[2\Phi(t,g)],\nonumber \\
&& \int d^3{\vec x} j_0^4(x) = \frac{g}{2}\times [h_{13}(t,g)+h^{*}_{13}(t,g)],\nonumber \\
&& \int d^3{\vec x} j_0^5(x) = -\frac{ig}{2}\times [h_{13}(t,g)-h^{*}_{13}(t,g)],\nonumber \\
&& \int d^3{\vec x} j_0^6(x) = \frac{g}{2}\times [h_{23}(t,g)+h^{*}_{23}(t,g)], \nonumber \\
&& \int d^3{\vec x} j_0^7(x) = -\frac{ig}{2}\times [h_{23}(t,g)-h^{*}_{23}(t,g)],\nonumber \\
&& \int d^3{\vec x} j_0^8(x) = \frac{g}{2\sqrt{3}}\times [1 -3~{\rm cos}^2\Theta(t,g)].
\label{q2shas}
\eea

\section{ General Form of Fundamental Color Charge of the Quark in Yang-Mills Theory in SU(3) }

Since the fundamental color charge vector ${\vec q}(t)$ is linearly proportional to $g$ (see eqs.
(\ref{ely}) and (\ref{qtg})) we find from eq. (\ref{q2shas}) that the color charge $q^a(t)$ of a
quark in Yang-Mills theory in SU(3) takes the form
\bea
&& q_1(t) = \frac{g}{2}\times [h_{12}(t)+h^{*}_{12}(t)], \nonumber \\
&& q_2(t)  = -\frac{ig}{2}\times[h_{12}(t)-h^{*}_{12}(t)],\nonumber \\
&& q_3(t)  = \frac{g}{2}\times{\rm sin}^2\Theta(t) \times{\rm cos}[2\Phi(t)]\nonumber \\
&& q_4(t)  = \frac{g}{2}\times [h_{13}(t)+h^{*}_{13}(t)],\nonumber \\
&& q_5(t)  = -\frac{ig}{2}\times [h_{13}(t)-h^{*}_{13}(t)],\nonumber \\
&& q_6(t)  = \frac{g}{2}\times [h_{23}(t)+h^{*}_{23}(t)], \nonumber \\
&& q_7(t)  = -\frac{ig}{2}\times [h_{23}(t)-h^{*}_{23}(t)],\nonumber \\
&& q_8(t)  = \frac{g}{2\sqrt{3}}\times [1 -3~{\rm cos}^2\Theta(t)].
\label{q2sh}
\eea
where the complex functions $h_{12}(t)$, $h_{13}(t)$, $h_{23}(t)$ and the real phases
$\Theta(t)$, $\Phi(t)$ depend on time $t$ but are independent of $g$. Since
$h_{12}(t)$, $h_{13}(t)$, $h_{23}(t)$ are complex functions we can write eq.
(\ref{q2sh}) as
\bea
&& q_1(t) = g\times |h_{12}(t)|\times {\rm cos}\phi_{12}(t) \nonumber \\
&& q_2(t) = g\times |h_{12}(t)|\times {\rm sin}\phi_{12}(t),\nonumber \\
&& q_3(t) = \frac{g}{2}\times{\rm sin}^2\Theta(t) \times{\rm cos}[2\Phi(t)], \nonumber \\
&& q_4(t) = g\times |h_{13}(t)|\times {\rm cos}\phi_{13}(t), \nonumber \\
&& q_5(t) = g\times |h_{13}(t)|\times {\rm sin}\phi_{13}(t), \nonumber \\
&& q_6(t) = g\times |h_{23}(t)|\times {\rm cos}\phi_{23}(t), \nonumber \\
&& q_7(t) = g\times |h_{23}(t)|\times {\rm sin}\phi_{23}(t), \nonumber \\
&& q_8(t) = \frac{g}{2\sqrt{3}}\times [1 -3~{\rm cos}^2\Theta(t)]
\label{qsp}
\eea
where the principal arguments
\bea
&& {\rm Arg}\left(h_{12}(t)\right)=\phi_{12}(t) = {\rm tan}^{-1}\left[\frac{{\rm Im}[h_{12}(t)]}{{\rm Re}[h_{12}(t)]}\right],~~~~~~~~~~{\rm Arg}\left(h_{13}(t)\right)=\phi_{13}(t) = {\rm tan}^{-1}\left[\frac{{\rm Im}[h_{13}(t)]}{{\rm Re}[h_{13}(t)]}\right],\nonumber \\
&&{\rm Arg}\left(h_{23}(t)\right)=\phi_{23}(t) = {\rm tan}^{-1}\left[\frac{{\rm Im}[h_{23}(t)]}{{\rm Re}[h_{23}(t)]}\right],
\label{qphio}
\eea
of the complex functions $h_{12}(t)$, $h_{13}(t)$ and $h_{23}(t)$ lying in the range
\bea
-\pi < \phi_{12}(t),~~ \phi_{13}(t),~~\phi_{23}(t) \le  \pi.
\label{qphin}
\eea

From eqs. (\ref{fly}) and (\ref{qsp}) we find
\bea
|h_{12}(t)|^2 + |h_{13}(t)|^2 +|h_{23}(t)|^2 +\left[\frac{1}{2} \times {\rm sin}^2\Theta(t) \times{\rm cos}[2\Phi(t)] \right]^2~+~\left[\frac{1}{2\sqrt{3}}\times [1-3~{\rm cos}^2\Theta(t)]\right]^2 =1.\nonumber \\
\label{qb2shj}
\eea
From eq. (\ref{qthtb}) we find that the maximum allowed range of $\Theta(t)$, $\Phi(t)$ are
\bea
0 \le \Theta(t) \le \pi,~~~~~~~~~~~~~~~~0 \le \Phi(t) < 2\pi.
\label{qthtbm}
\eea

In order to proceed further we need to find the maximum and minimum values of
\bea
&& H=\left[ {\rm sin}^2\Theta(t) \times{\rm cos}[2\Phi(t)] \right]^2~+~\left[\frac{1}{\sqrt{3}}\times [1-3~{\rm cos}^2\Theta(t)]\right]^2 \nonumber \\
&&=[{\rm cos}^2\Theta(t)-1]^2 \times{\rm cos}^2[2\Phi(t)]+\frac{1}{3}+~[3~{\rm cos}^4\Theta(t)-2~{\rm cos}^2\Theta(t)]
\label{h}
\eea
in the range of $\Theta(t)$, $\Phi(t)$ given by eq. (\ref{qthtbm}).
Taking the first derivative with respect to $\Phi(t)$ we find
\bea
\frac{dH}{d\Phi(t)}=-2~ {\rm sin}[4\Phi(t)]\times [{\rm cos}^2\Theta(t)-1]^2.
\label{bhda}
\eea
Hence we find
\bea
\frac{dH}{d\Phi(t)}=0
\label{bhd}
\eea
when
\bea
&&1)~{\rm cos}^2\Theta(t)=1,~~~~~~~~~~~{\rm for~any}~~~~~~~~\Phi(t);\nonumber \\
&&2)~\Phi(t)=0,~\frac{\pi}{4},~\frac{\pi}{2},~\frac{3\pi}{4},~\pi,~\frac{5\pi}{4},~\frac{3\pi}{2},~\frac{7\pi}{4},~~~~~~~~~~~{\rm for~any}~~~~~~~~\Theta(t).
\label{bhe}
\eea
Taking the first derivative with respect to $\Theta(t)$ we find
\bea
\frac{dH}{d\Theta(t)}=- 2~{\rm sin}[2\Theta(t)]\times \left[[{\rm cos}^2\Theta(t)-1]\times{\rm cos}^2[2\Phi(t)]+3~{\rm cos}^2\Theta(t)-1\right].
\label{bhb}
\eea
Hence we find
\bea
\frac{dH}{d\Theta(t)}=0=\frac{dH}{d\Phi(t)}
\label{both}
\eea
when
\bea
&&1)~{\rm cos}^2\Theta(t)=1,~~~~~~~~~~~{\rm for~any}~~~~~~~~\Phi(t);\nonumber \\
&&2)~\Phi(t)=0,~\frac{\pi}{2},~\pi,~\frac{3\pi}{2},~~~~~~~~~~~~~~~{\rm and}~~~~~~~~{\rm cos}^2\Theta(t)=\frac{1}{2}\nonumber \\
&&3)~\Phi(t)=\frac{\pi}{4},~\frac{3\pi}{4},~\frac{5\pi}{4},~\frac{7\pi}{4},~~~~~~~~~~~{\rm and}~~~~~~~~{\rm cos}^2\Theta(t)=\frac{1}{3}\nonumber \\
&&4)~\Phi(t)=0,~\frac{\pi}{2},~\pi,~\frac{3\pi}{2},~~~~~~~~~~~~~~~~{\rm and}~~~~~~~~\Theta(t)=\frac{\pi}{2}\nonumber \\
&&5)~\Phi(t)=\frac{\pi}{4},~\frac{3\pi}{4},~\frac{5\pi}{4},~\frac{7\pi}{4},~~~~~~~~~~~~{\rm and}~~~~~~~~\Theta(t)=\frac{\pi}{2}.
\label{1che}
\eea

By taking the second derivative we find
\bea
&& \frac{d^2H}{d\Theta^2(t)}=-4~{\rm cos}[2\Theta(t)]\times \left[[{\rm cos}^2\Theta(t)-1]\times{\rm cos}^2[2\Phi(t)]+3~{\rm cos}^2\Theta(t)-1\right] \nonumber \\
&&+2~{\rm sin}^2[2\Theta(t)]\times [{\rm cos}^2[2\Phi(t)]+3]
\label{bhf}
\eea
and
\bea
\frac{d^2H}{d\Phi^2(t)}=-8~ {\rm cos}[4\Phi(t)]\times [{\rm cos}^2\Theta(t)-1]^2
\label{bhg}
\eea
and
\bea
\frac{d^2H}{d\Theta(t)~d\Phi(t)}=4~ {\rm sin}[4\Phi(t)]\times {\rm sin}[2\Theta(t)]\times[{\rm cos}^2\Theta(t)-1].
\label{bhh}
\eea
We write
\bea
D= \frac{d^2H}{d\Theta^2(t)}\times \frac{d^2H}{d\Phi^2(t)} - [\frac{d^2H}{d\Theta(t)~d\Phi(t)}]^2.
\label{bhj}
\eea
\\

When
\bea
{\rm cos}^2\Theta(t)=1,~~~~~~~~~~~{\rm for~any}~~~~~~~~\Phi(t)
\label{c1}
\eea
we find
\bea
D = 0
\eea
which implies that the second derivative test is inconclusive.

When
\bea
\Phi(t)=0,~\frac{\pi}{2},~\pi,~\frac{3\pi}{2},~~~~~~~~~~~{\rm and}~~~~~~~~{\rm cos}^2\Theta(t)=\frac{1}{2}
\label{c2}
\eea
we find
\bea
D =- 16~{\rm sin}^2[2\Theta(t)] <0
\eea
which implies that it is saddle point.

When
\bea
\Phi(t)=\frac{\pi}{4},~\frac{3\pi}{4},~\frac{5\pi}{4},~\frac{7\pi}{4},~~~~~~~~~~~{\rm and}~~~~~~~~\Theta(t)=\frac{\pi}{2}
\label{c4}
\eea
we find
\bea
D =-32<0
\eea
which implies that it is saddle point.

When
\bea
\Phi(t)=\frac{\pi}{4},~\frac{3\pi}{4},~\frac{5\pi}{4},~\frac{7\pi}{4},~~~~~~~~~~~{\rm and}~~~~~~~~{\rm cos}^2\Theta(t)=\frac{1}{3}
\label{c3}
\eea
we find
\bea
D =\frac{64}{3}~{\rm sin}^2[2\Theta(t)] >0,~~~~~~~~~~~\frac{d^2H}{d\Theta^2(t)}=6~{\rm sin}^2[2\Theta(t)]>0
\eea
which gives the minimum value
\bea
H_{\rm min}=0.
\label{hmin}
\eea
\\

When
\bea
\Phi(t)=0,~\frac{\pi}{2},~\pi,~\frac{3\pi}{2},~~~~~~~~~~~{\rm and}~~~~~~~~\Theta(t)=\frac{\pi}{2}
\label{c5}
\eea
we find
\bea
D =64>0,~~~~~~~~~~~\frac{d^2H}{d\Theta^2(t)}=-8<0
\eea
which gives the maximum value
\bea
H_{\rm max}=\frac{4}{3}.
\label{hmax}
\eea

Hence we find from eqs. (\ref{h}), (\ref{hmin}) and (\ref{hmax}) that
\bea
0\le \left[\frac{1}{2} \times {\rm sin}^2\Theta(t) \times{\rm cos}[2\Phi(t)] \right]^2~+~\left[\frac{1}{2\sqrt{3}}\times [1-3~{\rm cos}^2\Theta(t)]\right]^2\le \frac{1}{3}.
\label{ch}
\eea
Using eq. (\ref{ch}) in eq. (\ref{qb2shj}) we find
\bea
\frac{2}{3}\le |h_{12}(t)|^2 + |h_{13}(t)|^2 +|h_{23}(t)|^2 \le 1.
\label{cst}
\eea
Since each individual square terms in eq. (\ref{qb2shj}) are positive we find from eqs. (\ref{qb2shj}), (\ref{ch})
and (\ref{cst}) that we can write
\bea
|h_{12}(t)|^2 + |h_{13}(t)|^2 +|h_{23}(t)|^2={\rm sin}^2\theta(t)
\label{hta}
\eea
and
\bea
\left[\frac{1}{2} \times {\rm sin}^2\Theta(t) \times{\rm cos}[2\Phi(t)] \right]^2~+~\left[\frac{1}{2\sqrt{3}}\times [1-3~{\rm cos}^2\Theta(t)]\right]^2 ={\rm cos}^2\theta(t)
\label{q2j}
\eea
where
\bea
\frac{2}{3}\le {\rm sin}^2\theta(t) \le 1.
\label{snt}
\eea
Since $|h_{12}(t)|$, $|h_{13}(t)|$ and $|h_{23}(t)|$ are positive
we write eq. (\ref{hta}) as
\bea
&& |h_{12}(t)| ={\rm sin}\theta(t) \times {\rm sin}\sigma(t) \times {\rm cos}\eta(t) \nonumber \\
&& |h_{13}(t)| ={\rm sin}\theta(t) \times {\rm sin}\sigma(t) \times {\rm sin}\eta(t) \nonumber \\
&& |h_{23}(t)| ={\rm sin}\theta(t) \times {\rm cos}\sigma(t)
\label{ht}
\eea
where
\bea
{\rm sin}^{-1}(\sqrt{\frac{2}{3}}) ~~\le ~~ \theta(t) ~~ \le ~~\pi-{\rm sin}^{-1}(\sqrt{\frac{2}{3}}),~~~~~~~~~~~0 \le \sigma(t),~\eta(t) \le \frac{\pi}{2}.
\label{asnt}
\eea
We write eqs. (\ref{q2j}) as
\bea
&& \frac{1}{2} \times {\rm sin}^2\Theta(t) \times{\rm cos}[2\Phi(t)] = {\rm cos}\theta(t) \times {\rm sin}\phi(t), \nonumber \\
&& \frac{1}{2\sqrt{3}}\times [1-3~{\rm cos}^2\Theta(t)] ={\rm cos}\theta(t) \times {\rm cos}\phi(t)
\label{aq2j}
\eea
where
\bea
0 \le \phi(t) \le 2\pi.
\label{bsnt}
\eea

From eqs. (\ref{ht}), (\ref{aq2j}), (\ref{qsp}), (\ref{qphin}), (\ref{asnt}) and (\ref{bsnt})
we find that the general form of eight time dependent fundamental color charges of the quark in
Yang-Mills theory in SU(3) is given by
\bea
&& q_1(t) = g\times {\rm sin}\theta(t) \times {\rm sin}\sigma(t)\times {\rm cos}\eta(t)\times {\rm cos}\phi_{12}(t), \nonumber \\
&& q_2(t) = g\times {\rm sin}\theta(t) \times {\rm sin}\sigma(t)\times {\rm cos}\eta(t)\times {\rm sin}\phi_{12}(t), \nonumber \\
&& q_3(t) = g\times {\rm cos}\theta(t) \times {\rm sin}\phi(t) \nonumber \\
&& q_4(t) = g\times {\rm sin}\theta(t) \times {\rm sin}\sigma(t)\times {\rm sin}\eta(t)\times {\rm cos}\phi_{13}(t), \nonumber \\
&& q_5(t) = g\times {\rm sin}\theta(t) \times {\rm sin}\sigma(t)\times {\rm sin}\eta(t)\times {\rm sin}\phi_{13}(t), \nonumber \\
&& q_6(t) = g\times {\rm sin}\theta(t)\times {\rm cos}\sigma(t) \times {\rm cos}\phi_{23}(t), \nonumber \\
&& q_7(t) = g\times {\rm sin}\theta(t)\times {\rm cos}\sigma(t)\times {\rm sin}\phi_{23}(t), \nonumber \\
&& q_8(t) = g\times {\rm cos}\theta(t)\times {\rm cos}\phi(t)
\label{qspn}
\eea
which reproduces eq. (\ref{qspin}) where
\bea
&&{\rm sin}^{-1}(\sqrt{\frac{2}{3}}) ~~\le ~~ \theta(t) ~~ \le ~~\pi-{\rm sin}^{-1}(\sqrt{\frac{2}{3}}),~~~~~~~~~~~0 \le \sigma(t),~\eta(t) \le \frac{\pi}{2},\nonumber \\
&& 0~\le ~ \phi(t)\le 2 \pi,~~~~~~~~~~~~~~~~~~~~~~~~~~~~~~~~-\pi < \phi_{12}(t),~\phi_{13}(t),~\phi_{23}(t) \le \pi.
\label{range}
\eea
Note that, as mentioned in the introduction, if all of these seven
phases become constants then all the eight color charges $q^a$ become constants in which case
the Yang-Mills potential $A^{\mu a}(x)$ reduces to Maxwell-like (abelian-like) potential
(see eqs. (\ref{fnab}) and (\ref{upg1fg}) or \cite{arxiv}). Since the abelian-like potential can not
explain confinement of quarks inside (stable) proton one finds that all the seven real phases
can not be constants. Hence one finds that
the general form of eight time dependent fundamental color charges $q^a(t)$ of the quark which we have
derived in eq. (\ref{qspn}) may provide an insight to the question why
quarks are confined inside a (stable) proton once the exact form of these time dependent phases
$\theta(t),~ \sigma(t),~\eta(t),~ \phi(t),~ \phi_{12}(t),~\phi_{13}(t),~\phi_{23}(t)$ are found out
[see section XVI for more discussion about this]. It should be remembered that the
static systems in the Yang-Mills theory are, in general, not
abelian-like. For example, unlike Maxwell theory where the electron at rest produces
(abelian) Coulomb potential, the quark at rest in Yang-Mills theory does not
produce abelian-like potential (see eq. (\ref{upg1fg}) and sections VII, VIII or see \cite{arxiv}).

\section{ Comparison with Wong's Equation }

In Maxwell theory the electromagnetic current density $j^\mu(x)$ of the electron
is linearly proportional to $e$. However, in Yang-Mills theory the situation is different.
The Yang-Mills potential $A^{\mu a}(x)$ in eq. (\ref{fnab}) contains infinite powers of
$g$ which implies that the Yang-Mills color current density
$j^{\mu a}(x)$ of the quark  in Yang-Mills theory in eq. (\ref{dcn}) [or in eq. (\ref{aqwn})]
contains infinite powers of $g$.

The Wong's equation \cite{wong}
\begin{equation}
\frac{dI^a(\tau)}{d\tau} = gf^{abc}A_\mu^c \frac{d\xi^\mu(\tau)}{d\tau} I^b(\tau)
\label{w1}
\end{equation}
of $I^a(\tau)$ with the world line $\xi^\mu(\tau)$ can be derived by using
\begin{equation}
j^{\mu a}(x) = g\int d\tau~ I^a(\tau) \frac{d\xi^\mu(\tau)}{d\tau} \delta^{(4)}(x-\xi(\tau))
\label{j1}
\end{equation}
in the equation
\begin{equation}
D_\nu[A]F^{\nu \mu a}(x)=j^{\mu a}(x)
\label{dfw}
\end{equation}
where $D_\mu^{ab}[A]$ is given by eq. (\ref{dab}) and $F^{\mu \nu a}(x)$ is given by
eq. (\ref{fmna}), see \cite{wong,stern,yashida}.

The Wong's equation [see eq. (\ref{w1})]
describes the precession of $I^a(\tau)$ in the background $A^{\mu a}$, which naturally contains all
orders in $g$  (unless $A^{\mu a}$ is of order
$\frac{1}{g}$) because the left hand side of eq. (\ref{w1}) is $\frac{dI^a(\tau)}{d\tau}$
and the right hand side of eq. (\ref{w1}) contains $gI^b(\tau)$. As mentioned above, the Yang-Mills potential
(color potential) $A^{\mu a}(x)$ produced by the quark contains
infinite powers of $g$ (see eq. (\ref{fnab}) or \cite{arxiv}) which implies that the Yang-Mills
color current density $j^{\mu a}(x)$
of the quark in Yang-Mills theory in eq. (\ref{dcn}) [or in eq. (\ref{aqwn})] contains infinite powers of $g$.
Hence from eqs. (\ref{dfw}) and (\ref{j1}) one finds that $I^a(t)$ contains infinite powers of $g$
which is consistent with eq. (\ref{w1}) because the left hand side of eq. (\ref{w1}) is $\frac{dI^a(\tau)}{d\tau}$
and the right hand side of eq. (\ref{w1}) contains $gI^b(\tau)$.  Since the definition of the fundamental time
dependent color charge $q^a(t)$ of the quark in eq. (\ref{qspn}) is linearly proportional to $g$ and the $I^a(t)$ in eq.
(\ref{j1}) [or in eq. (\ref{w1})] contains infinite powers of $g$, we find that the
fundamental time dependent color charge $q^a(t)$ of the quark in eq. (\ref{qspn}) is different
from the $I^a(t)$ used in the Wong's equation in \cite{wong}.

In our study we have followed the Yang-Mills paper \cite{yang} where the Yang-Mills
color current density $j^{\mu a}(x)$ of the quark as given by eqs. (\ref{dcn}) and (\ref{aqwn})
does not obey the continuity equation ($\partial_\mu j^{\mu a}(x)\neq 0$)
[but obeys the equation $D_\mu[A]j^{\mu a}(x)=0$, see eq. (\ref{ccc}), which implies that the color
charge $q^a(t)$ of the quark is time dependent] but the isotopic spin current density, say  ${\cal T}^{\mu a}(x)$,
of the system where
\bea
{\cal I}^{\mu a}(x)=j^{\mu a}(x)+gf^{abc}A_\nu^b(x)F^{\mu \nu c}(x)
\label{iscd}
\eea
(as defined in the Yang-Mills paper \cite{yang}) obeys the continuity equation
\bea
\partial_\mu {\bf {\cal I}}^{\mu a}(x)=0,
\label{icce}
\eea
(see eqs. (15) and (16) of \cite{yang}). Note that, as mentioned earlier,
the Yang-Mills theory was developed by making analogy with the corresponding procedure
in Maxwell theory. Hence, similar to the procedure in the Yang-Mills paper, we have followed
the analogy with the Maxwell theory to determine the general form of the fundamental time
dependent color charge $q^a(t)$ of the quark in eq. (\ref{qspn}).
In Maxwell theory the electron is a fundamental particle of
the nature. Hence in analogy to Maxwell theory one finds that the quark in Yang-Mills theory
is a fundamental particle of the nature. In Maxwell theory the electric charge $e$ of the
electron is a fundamental charge of the nature and hence it is independent of the Maxwell
potential $A^\mu(x)$. Hence in analogy to Maxwell theory one finds that since the color charge $q^a(t)$ of the
quark in Yang-Mills theory is a fundamental charge of the nature it is independent
of the Yang-Mills potential $A^{\mu a}(x)$. This can also be seen as follows.

Since the fundamental time dependent color charge $q^a(t)$ of the quark
in eq. (\ref{qspn}) is linearly proportional to $g$ and the Yang-Mills potential $A^{\mu a}(x)$ in
eq. (\ref{fnab}) [see also \cite{arxiv}] contains infinite powers of $g$ we find that the definition of the fundamental
time dependent color charge $q^a(t)$ of the quark in eq. (\ref{qspn}) is independent of the Yang-Mills
potential $A^{\mu a}(x)$.

\section{Second (Cubic) Casimir Invariant of SU(3) and General Form of Color Charges of the Quark}

Note that there is one Casimir invariant (quadratic Casimir invariant)
\bea
C=q_i(t)q_i(t),~~~~~~~~~~~~~~~~~~~~~~~~~~~~i=1,2,3
\label{cas}
\eea
of SU(2) which is gauge invariant with respect to the gauge transformation given by
eq. (\ref{r2}). We have fixed the exact value of the Casimir invariant $q_i(t)q_i(t)=g^2$ of SU(2)
[see eqs. (\ref{b2shj}) and (\ref{gly})] in the derivation of the general form of three time dependent
color charges $q_i(t)$ of a fermion in SU(2) in eq.  (\ref{2skfn}) where $i=1,2,3$.
Since we have fixed the exact value of the Casimir invariant $q_i(t)q_i(t)=g^2$ of SU(2)
[see eqs. (\ref{b2shj}) and (\ref{gly})] we find that the general form of
three time dependent color charges $q_1(t),~q_2(t),~q_3(t)$ of a fermion in SU(2) in eq.  (\ref{2skfn})
depend on $g$ and two time dependent phases $\theta(t),~\phi(t)$. It can be
observed that the relation between three time dependent color charges $q_1(t),~q_2(t),~q_3(t)$
of a fermion in SU(2) and $g,~\theta(t),~\phi(t)$ which we have found in eq. (\ref{2skfn}) is similar to the relation
between cartesian coordinates $x_1,x_2,x_3$ and spherical polar coordinates $r,\theta_1,\theta_2$
in three dimensions where
\bea
&& x_1 =r \times {\rm sin}\theta_1\times {\rm cos}\theta_2,\nonumber \\
&& x_2 = r\times {\rm sin}\theta_1\times {\rm sin}\theta_2,\nonumber \\
&& x_3 = r\times {\rm cos}\theta_1, \nonumber \\
&&0\le \theta_1 \le \pi,~~~~~~~~0 \le \theta_2 < 2\pi,~~~~~~r^2=x_ix_i
\label{3dr}
\eea
[except that the ranges of time dependent phases in eq. (\ref{2skfn}) are different, see eq. (\ref{2fna}),
which is due the fact that $\tau^i=\frac{\sigma^i}{2}$ in SU(2) in eq. (\ref{paulim})].
The time dependent gauge transformation of the
color charge $q_i(t)$ of a fermion in the adjoint representation of SU(2) in eq. (\ref{r2})
corresponds to a time dependent rotation in SO(3) [see section IX for more discussion on this].
The general form of three time dependent color charges $q_i(t)$ of a fermion in SU(2) in eq.
(\ref{2skfn}) is consistent with the fact that
the SU(2) and SO(3) are locally isomorphic.

The situation is different in SU(3) as we will see below.

In SU(3) there are two independent Casimir invariants
\bea
C_1=q^a(t)q^a(t),~~~~~~~~~~~~~~~~~~~~~~~~~~~~~~~~~~~~a=1,2,...,8
\label{1cas}
\eea
and
\bea
C_2=d_{abc}q^a(t)q^b(t)q^c(t),~~~~~~~~~~~~~~~~~~~~~~~~~~~~~~~a,b,c=1,2,...,8
\label{2cas}
\eea
where $d_{abc}$ are the symmetric structure constants of the SU(3) group. Both the Casimir invariants
of SU(3) in eqs. (\ref{1cas}) and (\ref{2cas}) are gauge invariant with respect to gauge transformation of the
color charge $q^a(t)$ of the quark as given by eq. (\ref{r3}).
As mentioned in section IX, the gauge
transformation in the adjoint representation of SU(3) in eq. (\ref{r3}) which is described by 8 real parameters
does not correspond to a time dependent general rotation in SO(8)
because a general rotation in SO(8) is not described by 8 real parameters but
a general rotation in SO(8) is described by 28 real parameters, see for example \cite{yeh}.
Hence one finds that for a general
vector $w^a$, since all the non-zero values of $d_{abc}$ are not same for different values of $a,b,c$ where
$a,b,c=1,2,...,8$, the $d_{abc}w^aw^bw^c$ is not rotationally invariant with respect to the general rotation
in SO(8) although $w^aw^a$ is rotationally invariant with respect to the general rotation
in SO(8).

Similarly, it can also be easily verified that since all the non-zero
values of $d_{abc}$ are not same for different values of $a,b,c$ where $a,b,c=1,2,...,8$
one finds that once the exact value of the first Casimir invariant (quadratic Casimir invariant)
\bea
q^a(t)q^a(t)=g^2
\label{qaqa}
\eea
of SU(3) is fixed, the second Casimir invariant (cubic Casimir invariant)
$C_2=d_{abc}q^a(t)q^b(t)q^c(t)$ of SU(3) satisfies the range \cite{rest1,rest1a,rest1b,so8}
\begin{equation}
-\frac{g^3}{\sqrt{3}}\le d_{abc}q^a(t)q^b(t)q^c(t) \le \frac{g^3}{\sqrt{3}}.
\label{dabc}
\end{equation}

Note that in the derivation of the general form of eight time dependent fundamental color charges $q^a(t)$
of the quark in eq. (\ref{qspn}) we have fixed the exact value of the first casimir invariant
$C_1=q^a(t)q^a(t)=g^2$ of SU(3) [see eqs. (\ref{qb2shj}) and (\ref{fly})] but we have not
fixed the exact value of the second casimir invariant (cubic casimir invariant)
$C_2=d_{abc}q^a(t)q^b(t)q^c(t)$ of SU(3) because it satisfies the range given by eq. (\ref{dabc}).
Since we have only fixed the exact value of the first casimir invariant
$C_1=q^a(t)q^a(t)=g^2$ of SU(3) [see eqs. (\ref{qb2shj}) and (\ref{fly})], the general form of eight fundamental
time dependent color charges $q_1(t),~q_2(t),~q_3(t),~q_4(t),~q_5(t),~q_6(t),~q_7(t),~q_8(t)$ of the quark in eq.
(\ref{qspn}) depend on $g$ and seven time dependent phases
$\theta(t),~ \sigma(t),~\eta(t),~ \phi(t),~ \phi_{12}(t),~\phi_{13}(t),~\phi_{23}(t)$.
However, unlike SU(2) case in eq. (\ref{2skfn}) which is similar to
spherical polar coordinates in three dimensions [see eq. (\ref{3dr})], the relation between
eight time dependent fundamental color charges $q_1(t),~q_2(t),~q_3(t),~q_4(t),~q_5(t),~q_6(t),~q_7(t),~q_8(t)$
of the quark in SU(3) and $g,~\theta(t),~ \sigma(t),~\eta(t),~ \phi(t),~ \phi_{12}(t),~\phi_{13}(t),~\phi_{23}(t)$
which we have found in eq. (\ref{qspn}) is not similar to the relation between cartesian coordinates
$x_1,x_2,x_3,x_4,x_5,x_6,x_7,x_8$ and spherical polar coordinates
$r,\theta_1,\theta_2,\theta_3,\theta_4,\theta_5,\theta_6,\theta_7$ in eight dimensions where
\bea
&& x_8=r\times {\rm sin}\theta_1 \times  {\rm sin}\theta_2 \times {\rm sin}\theta_3 \times {\rm sin}\theta_4 \times {\rm sin}\theta_5 \times {\rm sin}\theta_6 \times {\rm sin}\theta_7, \nonumber \\
&& x_7=r\times {\rm sin}\theta_1 \times  {\rm sin}\theta_2 \times {\rm sin}\theta_3 \times {\rm sin}\theta_4 \times {\rm sin}\theta_5 \times {\rm sin}\theta_6 \times {\rm cos}\theta_7, \nonumber \\
&& x_6=r\times {\rm sin}\theta_1 \times  {\rm sin}\theta_2 \times {\rm sin}\theta_3 \times {\rm sin}\theta_4 \times {\rm sin}\theta_5 \times {\rm cos}\theta_6, \nonumber \\
&& x_5=r\times {\rm sin}\theta_1 \times  {\rm sin}\theta_2 \times {\rm sin}\theta_3 \times {\rm sin}\theta_4 \times {\rm cos}\theta_5, \nonumber \\
&& x_4=r\times {\rm sin}\theta_1 \times  {\rm sin}\theta_2 \times {\rm sin}\theta_3 \times {\rm cos}\theta_4, \nonumber \\
&& x_3=r\times {\rm sin}\theta_1 \times  {\rm sin}\theta_2 \times {\rm cos}\theta_3, \nonumber \\
&& x_2=r\times {\rm sin}\theta_1 \times  {\rm cos}\theta_2, \nonumber \\
&& x_1=r\times {\rm cos}\theta_1, \nonumber \\
&& 0\le \theta_1, \theta_2, \theta_3, \theta_4, \theta_5, \theta_6 \le \pi,~~~~0 \le \theta_7 <  2\pi,~~~~~~r^2=x_ax_a
\label{8dr}
\eea
even if we have only fixed the exact value of the first casimir invariant $C_1=q^a(t)q^a(t)=g^2$ of SU(3)
[see eqs. (\ref{qb2shj}) and (\ref{fly})] to derive eq. (\ref{qspn}). Hence the general form of eight
time dependent color charges $q^a(t)$ of the quark which we have found in eq. (\ref{qspn})
reflect the fact that there is a second casimir invariant (cubic casimir invariant)
$C_2=d_{abc}q^a(t)q^b(t)q^c(t)$ of SU(3) whose range is given by eq. (\ref{dabc}).
One may wonder that since we have not fixed the exact value of the
second casimir invariant (cubic casimir invariant)
$C_2=d_{abc}q^a(t)q^b(t)q^c(t)$ of SU(3) in the derivation of eq. (\ref{qspn})
how can the the general form of eight time dependent fundamental color charges $q^a(t)$
of the quark in eq. (\ref{qspn}) reflect the fact that there is a second casimir
invariant (cubic casimir invariant) $C_2=d_{abc}q^a(t)q^b(t)q^c(t)$ of SU(3).
The answer to this question is that we have used the exact expressions of eight
generators $T^a$ of SU(3) by using exact expressions of eight Gell-Mann matrices
from eq. (\ref{T}) in the color current density $j^{\mu a}(x)$ of the quark
in eq. (\ref{q2sa}) to derive eq. (\ref{qspn}) which makes it implicit that there is a second
casimir invariant (cubic casimir invariant) $C_2=d_{abc}q^a(t)q^b(t)q^c(t)$ of SU(3) even if
we have not fixed its exact value because it satisfies the range given by eq. (\ref{dabc}).
One way to see this is to observe the corresponding situation in SU(2). In SU(2) we have used exact expressions of
three generators $\tau^i$ of SU(2) by using exact expressions of three Pauli matrices
from eq. (\ref{paulim}) in the color current density $j^{\mu i}(x)$ of a fermion
in SU(2) in eq. (\ref{2sa}) to derive eq. (\ref{2skfn}) which makes it implicit that the
cubic casimir invariant is absent in SU(2) and hence the
general form of three time dependent color charges $q_1(t),~q_2(t),~q_3(t)$ of a fermion in SU(2)
in eq. (\ref{2skfn}) is found to be similar to spherical polar coordinates in three dimensions.
Hence we find that the general form of eight time dependent fundamental color charges $q^a(t)$
of the quark in eq. (\ref{qspn}) is consistent with the fact that there is a second casimir
invariant (cubic casimir invariant) $C_2=d_{abc}q^a(t)q^b(t)q^c(t)$ of SU(3) but its exact
value is not fixed because it satisfies the range given by eq. (\ref{dabc}).
As mentioned above the time dependent gauge transformation of the color charge $q^a(t)$ of the quark
in the adjoint representation of SU(3) in eq. (\ref{r3})
does not correspond to a time dependent general rotation in SO(8).
Hence we find that in the Yang-Mills theory the general form of three time dependent color
charges $q_i(t)$ of a fermion in SU(2) in eq. (\ref{2skfn}) and the general form of eight time
dependent fundamental color charges $q^a(t)$ of the quark in SU(3) in eq.
(\ref{qspn}) are consistent with the fact that SU(2) and SO(3) are locally isomorphic, while
SU(3) and SO(8) are not.

In order to provide an expression for how the second Casimir invariant (cubic Casimir invariant)
$C_2=d_{abc}q^a(t)q^b(t)q^c(t)$ of SU(3) depends on seven time dependent phases
$\theta(t),~ \sigma(t),~\eta(t),~ \phi(t),~ \phi_{12}(t),~\phi_{13}(t),~\phi_{23}(t)$ of the
color charges $q^a(t)$ of the quark [see eq. (\ref{qspn})] we proceed as follows. From eq. (\ref{T}) we find
\bea
&&T^aq^a(t) =\frac{1}{2} \left(\begin{array}{ccc}
q_3(t)+\frac{q_8(t)}{\sqrt{3}},&q_1(t)-iq_2(t),&q_4(t)-iq_5(t)\\
q_1(t)+iq_2(t),&-q_3(t)+\frac{q_8(t)}{\sqrt{3}},&q_6(t)-iq_7(t)\\
q_4(t)+iq_5(t),&q_6(t)+iq_7(t),&-\frac{2q_8(t)}{\sqrt{3}}
\end{array} \right)
\label{Tdet}
\eea
which gives
\bea
&&{\rm Det}[T^aq^a(t)] =\frac{q_8(t)}{8\sqrt{3}}[3q^2_3(t)+\frac{q^2_8(t)}{3}-g^2]
\nonumber \\
&&+[q^2_1(t)+q^2_2(t)]\frac{3q_8(t)}{8\sqrt{3}}-\frac{q_3(t)}{8}[q^2_6(t)+q^2_7(t)]+\frac{q_3(t)}{8}[q^2_4(t)+q^2_5(t)]\nonumber \\
&&+\frac{q_1(t)}{4}[q_4(t)q_6(t)+q_5(t)q_7(t)]+\frac{q_2(t)}{4}[q_5(t)q_6(t)-q_4(t)q_7(t)]
\label{2det}
\eea
where $g^2$ is given by eq. (\ref{fly}). Since \cite{rest1a}
\bea
{\rm Det}[T^aq^a(t)]=\frac{1}{12}d_{abc}q^a(t)q^b(t)q^c(t)
\label{1det}
\eea
we find from eqs. (\ref{1det}) and (\ref{2det}) that
\bea
&&d_{abc}q^a(t)q^b(t)q^c(t)=\frac{3q_8(t)}{2\sqrt{3}}[3q^2_3(t)+\frac{q^2_8(t)}{3}-g^2]+\frac{3\sqrt{3}q_8(t)}{2}[q^2_1(t)+q^2_2(t)]
\nonumber \\
&&+\frac{3q_3(t)}{2}[q^2_4(t)+q^2_5(t)]-\frac{3q_3(t)}{2}[q^2_6(t)+q^2_7(t)]\nonumber \\
&&+3q_1(t)[q_4(t)q_6(t)+q_5(t)q_7(t)]+3q_2(t)[q_5(t)q_6(t)-q_4(t)q_7(t)].
\label{3det}
\eea
The eq. (\ref{3det}) can also be verified by directly using the non-zero symmetric structure constants
$d_{abc}$ of SU(3) group
\bea
\begin{array}{ccccccccccccccccc}
{\underline{abc}}~&118~&146~&157~&228~&247~&256~&338~&344~&355
~&366~&377~&448~&558~&668~&778~&888\\
{\underline {d_{abc}}}~&\frac{1}{\sqrt{3}}~&\frac{1}{2}~&\frac{1}{2}~&\frac{1}{\sqrt{3}}~&-\frac{1}{2}~&\frac{1}{2}~&\frac{1}{\sqrt{3}}~&\frac{1}{2}~&\frac{1}{2}
~&-\frac{1}{2}~&-\frac{1}{2}~&-\frac{1}{2\sqrt{3}}~&-\frac{1}{2\sqrt{3}}~&-\frac{1}{2\sqrt{3}}~&-\frac{1}{2\sqrt{3}}~&-\frac{1}{\sqrt{3}}
\end{array}\nonumber \\
\label{dabclist}
\eea
which gives
\bea
&&d_{abc}q^a(t)q^b(t)q^c(t)=3[d_{118}q_1(t)q_1(t)q_8(t)+d_{228}q_2(t)q_2(t)q_8(t)+d_{338}q_3(t)q_3(t)q_8(t)\nonumber \\
&&+d_{344}q_3(t)q_4(t)q_4(t)+
d_{355}q_3(t)q_5(t)q_5(t)+d_{366}q_3(t)q_6(t)q_6(t)+d_{377}q_3(t)q_7(t)q_7(t)\nonumber \\
&&+d_{448}q_4(t)q_4(t)q_8(t)+d_{558}q_5(t)q_5(t)q_8(t)+d_{668}q_6(t)q_6(t)q_8(t)+d_{778}q_7(t)q_7(t)q_8(t)]\nonumber \\
&&+6[d_{146}q_1(t)q_4(t)q_6(t)+d_{157}q_1(t)q_5(t)q_7(t)+d_{247}q_2(t)q_4(t)q_7(t)+d_{256}q_2(t)q_5(t)q_6(t)]\nonumber \\
&&+d_{888}q_8(t)q_8(t)q_8(t).
\label{4det}
\eea
Using the non-zero values of symmetric structure constants $d_{abc}$ of SU(3) from eq. (\ref{dabclist}) in eq. (\ref{4det}) we find
\bea
&&d_{abc}q^a(t)q^b(t)q^c(t)=\frac{q^3_8(t)}{2\sqrt{3}}+\frac{9}{2\sqrt{3}}q^2_3(t)q_8(t)+\frac{9}{2\sqrt{3}}q_8(t)[q^2_1(t)+q^2_2(t)]\nonumber \\
&&+3q_1(t)[q_4(t)q_6(t)+q_5(t)q_7(t)]+3q_2(t)[q_5(t)q_6(t)-q_4(t)q_7(t)]\nonumber \\
&&+\frac{3q_3(t)}{2}[q^2_4(t)+q^2_5(t)]-\frac{3q_3(t)}{2}[q^2_6(t)+q^2_7(t)]-\frac{3q_8(t)}{2\sqrt{3}}g^2
\eea
which reproduces eq. (\ref{3det}) where $g^2$ is given by eq. (\ref{fly}).

From eqs. (\ref{qspn}) and (\ref{3det}) we find that the expression of the second Casimir invariant
(cubic Casimir invariant) $C_2=d_{abc}q^a(t)q^b(t)q^c(t)$ of SU(3) in terms of seven time dependent phases
$\theta(t),~ \sigma(t),~\eta(t),~ \phi(t),~ \phi_{12}(t),~\phi_{13}(t),~\phi_{23}(t)$ of the
color charges $q^a(t)$ of the quark [see eq. (\ref{qspn})] is given by
\bea
&&d_{abc}q^a(t)q^b(t)q^c(t)= \frac{3 g^3}{2}{\rm sin}^3\theta(t) ~ {\rm sin^2}\sigma(t)~{\rm cos}\sigma(t)~ {\rm sin}2\eta(t)~ {\rm cos}[\phi_{12}(t)-\phi_{13}(t)+\phi_{23}(t)]\nonumber \\
&&+\frac{\sqrt{3}g^3}{2} {\rm cos}\theta(t) ~{\rm cos}\phi(t)~ [3~{\rm sin}^2\theta(t)~ {\rm sin}^2\sigma(t) ~{\rm cos}^2\eta(t)+3~{\rm cos}^2\theta(t)~ {\rm sin}^2\phi(t)+\frac{{\rm cos}^2\theta(t)~{\rm cos}^2\phi(t)}{3}-1]\nonumber \\
&&+\frac{3g^3}{2}~ {\rm cos}\theta(t) ~ {\rm sin}\phi(t)~ {\rm sin}^2\theta(t) ~ [{\rm sin}^2\sigma(t)~ {\rm sin}^2\eta(t)- {\rm cos}^2\sigma(t)].
\label{dabcp}
\eea

Note that the universal coupling $g$
(which is the physical observable, a fundamental quantity
of the nature) is the only parameter (apart from the mass $m$ of the quark)
that appears in the classical Yang-Mills lagrangian density
\cite{yang,muta} (see eq. (\ref{yml}) above).
As mentioned above, from eqs. (\ref{fly}) and (\ref{qb2shj})
one finds that the first Casimir (quadratic Casimir) invariant $C_1=q^a(t)q^a(t)=g^2$ of SU(3)
is fixed to be $g^2$ which is a physical observable. Hence we find that
the general form of eight time dependent fundamental color charges
$q^a(t)$ of the quark in Yang-Mills theory in SU(3) in eq. (\ref{qspn})
depend on $g$ and seven time dependent phases
$\theta(t),~ \sigma(t),~\eta(t),~ \phi(t),~ \phi_{12}(t),~\phi_{13}(t),~\phi_{23}(t)$
where the ranges of these seven time dependent phases are given by eq. (\ref{range}).
Since the first Casimir (quadratic Casimir) invariant $q^a(t)q^a(t)$
and the second Casimir (cubic Casimir) invariant $d_{abc}q^a(t)q^b(t)q^c(t)$ of SU(3) are two
independent Casimir invariants, one expects that if the second Casimir (cubic Casimir) invariant
$d_{abc}q^a(t)q^b(t)q^c(t)$ of SU(3)
corresponds to any physical observable then that physical observable should be experimentally measured
and that physical observable should be different from $g$ because the first Casimir (quadratic Casimir)
invariant $q^a(t)q^a(t)$ of SU(3) is fixed to be $g^2$, see eqs. (\ref{fly}) and (\ref{qb2shj}).
If such a physical observable exists in the nature and is fixed to be, say $C_3$,
[for example by experiments] where the fixed $C_3$ is given by eq. (\ref{3fix})
then one finds from eqs. (\ref{dabcp}) and (\ref{3fix}) that
\bea
&&\phi_{13}(t)~=~\phi_{12}(t)+\phi_{23}(t)~+~{\rm cos}^{-1}[\frac{1}{{\rm sin}^3\theta(t) ~ {\rm sin^2}\sigma(t)~{\rm cos}\sigma(t)~ {\rm sin}2\eta(t)}\times [\frac{2C_3}{3g^3}\nonumber \\
&&-~ {\rm sin}^2\theta(t)~ {\rm cos}\theta(t) ~ {\rm sin}\phi(t) ~ [{\rm sin}^2\sigma(t)~ {\rm sin}^2\eta(t)- {\rm cos}^2\sigma(t)]\nonumber \\
&&-\frac{1}{\sqrt{3}} ~{\rm cos}\theta(t) ~{\rm cos}\phi(t)~ [3~{\rm sin}^2\theta(t)~ {\rm sin}^2\sigma(t) ~{\rm cos}^2\eta(t)+3~{\rm cos}^2\theta(t)~ {\rm sin}^2\phi(t)+\frac{{\rm cos}^2\theta(t)~{\rm cos}^2\phi(t)}{3}-1]]]\nonumber \\
\label{phi13}
\eea
in which case the general form of eight time dependent fundamental
color charges $q^a(t)$ of the quark in eq. (\ref{qspn}) depend on $g$, $C_3$ and six time dependent phases
$\theta(t),~ \sigma(t),~\eta(t),~ \phi(t),~ \phi_{12}(t),~\phi_{23}(t)$ where $\phi_{13}(t)$ is
given by eq. (\ref{phi13}). However, note that even if all the physical observables are gauge invariant
but not all the gauge invariants are physical observables. Hence if there exists no physical
observable in the nature which is related to the fixed value $C_3$ as given by eq. (\ref{3fix})
[for example if one can not find any such observable from the experiments] then the second Casimir invariant
(cubic Casimir invariant) $d_{abc}q^a(t)q^b(t)q^c(t)$ of SU(3) satisfies the range
$-\frac{g^3}{\sqrt{3}}\le d_{abc}q^a(t)q^b(t)q^c(t) \le \frac{g^3}{\sqrt{3}}$
[see eq. (\ref{dabc})] in which case the general form of eight time dependent fundamental
color charges $q^a(t)$ of the quark in eq. (\ref{qspn}) depend on $g$ and seven time dependent phases
$\theta(t),~ \sigma(t),~\eta(t),~ \phi(t),~ \phi_{12}(t),~\phi_{13}(t),~\phi_{23}(t)$
where the ranges of these seven time dependent phases are given by eq. (\ref{range}).
Hence we find that the general form of eight time dependent fundamental color charges of the quark
in Yang-Mills theory in SU(3) is given by eq. (\ref{qspn}) where
$\theta(t),~\sigma(t),~\eta(t),~\phi(t),~\phi_{12}(t),~\phi_{13}(t),~\phi_{23}(t)$
are real time dependent phases.

\section{ Advantage of Time Dependent Phases in the Color Charge of the Quark }

One of the main difference between Maxwell theory and Yang-Mills theory is that
while Maxwell potential $A^\mu(x)$, Maxwell field tensor $F^{\mu \nu}(x)$,
Maxwell electric current density $j^\mu(x)$ of the electron
are linearly proportional to the electron charge $e$; the non-abelian Yang-Mills potential $A^{\mu a}(x)$,
the non-abelian Yang-Mills field tensor $F^{\mu \nu a}(x)$, the non-abelian
Yang-Mills color current density $j^{\mu a}(x)$ of the quark contain infinite
powers of $g$ (see eqs. (\ref{fnab}), (\ref{fmna}), and (\ref{dcn})).

Hence one finds that the classical Maxwell theory is in the perturbative regime of the quantum
electrodynamics (QED), whereas the classical Yang-Mills theory is in the non-perturbative regime
of the quantum chromodynamics (QCD). Hence it may not be surprising if the Yang-Mills potential $A^{\mu a}(x)$
produced by the quark may provide an
insight to the question why quarks are confined inside a (stable) proton, similar to Coulomb
potential $A_0(x)$ which explains (stable) hydrogen atom in Bohr's atomic model.

The general form of eight time dependent fundamental color charges $q^a(t)$ of the quark derived in this paper
depend on the strong-coupling constant $g$ and time dependent phases
$\theta(t),~ \sigma(t),~\eta(t),~ \phi(t),~ \phi_{12}(t),~\phi_{13}(t),~\phi_{23}(t)$
[see eq. (\ref{qspn})]. When all these phases become constants then the eight color charges
$q^a$ become constants in which case the Yang-Mills potential $A^{\mu a}(x)$ in eq. (\ref{fnab}) reduces to Maxwell-like
(abelian-like) potential (see also eq. (\ref{upg1fg}) and sections VII, VIII or \cite{arxiv}).
Since abelian-like potential can not explain confinement of quarks inside (stable) proton, one finds
that all these phases can not be constants.
The information about confinement of quarks inside (stable) proton can be obtained from the exact form of
the gauge invariant $F_{\mu \nu }^a(x)F^{\mu \nu a}(x)$. However, the exact form of the gauge invariant
$F_{\mu \nu}^a(x)F^{\mu \nu a}(x)$ can not be obtained from the gauge invariant $q^a(t)q^a(t)={\vec q}^2(t)=g^2$, but the exact
form of the gauge invariant $F_{\mu \nu}^a(x)F^{\mu \nu a}(x)$
can be obtained [see eq. (\ref{fmna})] from the exact form of Yang-Mills potential $A^{\mu a}(x)$
which can be obtained from the exact form of the color charges $q^a(t)$ of the quark, see eq. (\ref{fnab}).
Hence in order to get an insight about confinement of quarks inside (stable) proton it is not enough to know the gauge
invariant $q^a(t)q^a(t)={\vec q}^2(t)=g^2$ but it is necessary to know the exact form of
the color charges $q^a(t)$ of the quark which means
we need to find out the exact form of these time dependent phases
$\theta(t),~ \sigma(t),~\eta(t),~ \phi(t),~ \phi_{12}(t),~\phi_{13}(t),~\phi_{23}(t)$.
Hence the advantage of the general form of the color charge $q^a(t)$
of the quark derived in this paper is that it may provide an
insight to the question why quarks are confined inside a (stable) proton
once the exact form of these time dependent phases
$\theta(t),~ \sigma(t),~\eta(t),~ \phi(t),~ \phi_{12}(t),~\phi_{13}(t),~\phi_{23}(t)$ are found out.

\section{Conclusion}

In Maxwell theory the constant electric charge $e$ of the electron is consistent with the continuity
equation $\partial_\mu j^\mu(x)=0$ where $j^\mu(x)$ is the current density of the electron. However,
in Yang-Mills theory the Yang-Mills color current density $j^{\mu a}(x)$ of the quark satisfies the equation
$D_\mu[A]j^{\mu a}(x)=0$ which is not a continuity equation ($\partial_\mu j^{\mu a}(x)\neq 0$)
which implies that the color charge of the quark is not constant where $a=1,2,...,8$ are the
color indices. Since the charge of a point particle is obtained from the zero ($\mu =0$) component of
a corresponding current density by integrating over the entire (physically) allowed volume, the
color charge $q^a(t)$ of the quark in Yang-Mills theory is time dependent. In this paper we have
derived the general form of eight time dependent fundamental color charges $q^a(t)$ of the quark
in Yang-Mills theory in SU(3) where $a=1,2,...,8$.
\acknowledgments

I thank George Sterman for useful discussions and suggestions.

\end{document}